\documentclass[sigconf, screen]{acmart}

\renewcommand\footnotetextcopyrightpermission[1]{} % Remove footnote space

\definecolor{wellesleyblue}{RGB}{0, 39, 118}

\newcommand{\theneu}[0]{Northeastern University}
\newcommand{\charlie}[0]{Nguyen et al. }

\copyrightyear{2025}
\acmYear{2025}
\setcopyright{cc}
\setcctype[4.0]{by}
\acmConference[FSE Companion '25]{33rd ACM International Conference on the Foundations of Software Engineering}{June 23--28, 2025}{Trondheim, Norway}
\acmBooktitle{33rd ACM International Conference on the Foundations of Software Engineering (FSE Companion '25), June 23--28, 2025, Trondheim, Norway}\acmDOI{10.1145/3696630.3731663}
\acmISBN{979-8-4007-1276-0/2025/06}
\acmDOI{10.1145/3696630.3731663}

\usepackage{ccicons}
\usepackage{wasysym}
\usepackage{subcaption}
\usepackage{listings}
\usepackage{multirow}
\usepackage{makecell}
\usepackage{array}
\usepackage{balance}

\begin{document}

\title{``I Would Have Written My Code Differently'': Beginners Struggle to Understand LLM-Generated Code}

\author{Yangtian Zi}
\affiliation{%
  \institution{Northeastern University}
  \city{Boston}
  \country{United States}
}
\email{zi.ya@northeastern.edu}

\author{Luisa Li}
\affiliation{%
  \institution{Northeastern University}
  \city{Boston}
  \country{United States}
}
\email{li.tao@northeastern.edu}

\author{Arjun Guha}
\affiliation{%
 \institution{Northeastern University}
  \city{Boston}
  \country{United States}
}
\email{a.guha@northeastern.edu}

\author{Carolyn~Jane Anderson}
\affiliation{%
 \institution{Wellesley College}
  \city{Wellesley}
  \country{United States}
}
\email{carolyn.anderson@wellesley.edu}

\author{Molly Q Feldman}
\affiliation{%
 \institution{Oberlin College}
  \city{Oberlin}
  \country{United States}
}
\email{mfeldman@oberlin.edu}

% \setcopyright{none}
% \settopmatter{printacmref=false}
% \renewcommand\footnotetextcopyrightpermission[1]{}

\begin{abstract}
% note: keep in mind that it needs to be within 150 words.

Large language models (LLMs) are being increasingly adopted for programming work. Prior work shows that while LLMs accelerate task completion for professional programmers, beginning programmers struggle to prompt models effectively. However, prompting is just half of the code generation process-- when code is generated, it must be read, evaluated, and integrated (or rejected). How accessible are these tasks for beginning programmers? 

This paper measures how well beginners comprehend LLM-generated code and explores the challenges students face in judging code correctness. We compare how well students understand natural language descriptions of functions and LLM-generated implementations, studying 32 CS1 students on 160 task instances. Our results show a low per-task success rate of 32.5\%, with indiscriminate struggles across demographic populations. Key challenges include barriers for non-native English speakers, unfamiliarity with Python syntax, and automation bias. Our findings highlight the barrier that code comprehension presents to beginning programmers seeking to write code with LLMs.
\end{abstract}

\begin{CCSXML}
<ccs2012>
<concept>
<concept_id>10003456.10003457.10003527</concept_id>
<concept_desc>Social and professional topics~Computing education</concept_desc>
<concept_significance>500</concept_significance>
</concept>
<concept>
<concept_id>10011007</concept_id>
<concept_desc>Software and its engineering</concept_desc>
<concept_significance>300</concept_significance>
</concept>
<concept>
<concept_id>10003120.10003121.10011748</concept_id>
<concept_desc>Human-centered computing~Empirical studies in HCI</concept_desc>
<concept_significance>500</concept_significance>
</concept>
</ccs2012>
\end{CCSXML}

\ccsdesc[500]{Social and professional topics~Computing education}
\ccsdesc[300]{Software and its engineering}
\ccsdesc[500]{Human-centered computing~Empirical studies in HCI}
\keywords{Large Language Models, Code Comprehension, Computer Science Education, CS1}


\maketitle

\begingroup
\renewcommand\thefootnote{}\footnotetext{\textit{FSE Companion '25, June 23--28, 2025, Trondheim, Norway}\\
© 2025 Copyright help by the owner/author(s).\\
\ccby{} This work is licensed under a Creative Commons Attribution International 4.0 License.\\
This is the author’s version of the work. It is posted here for your personal use. Not for redistribution.\\
The definitive Version of Record was published in Proceedings of \textit{the 33rd ACM International Conference on the Foundations of Software Engineering (FSE Companion '25), June 23--28, 2025, Trondheim, Norway.}}
\endgroup

\section{Introduction}

Today, large language models (LLMs) are widely used by professional programmers to accelerate their work. 
For example, Google recently reported that more than a quarter of new code at Google is AI-generated, then reviewed and accepted by engineers~\citep{Q3EarningsCall2024}, with other major software companies making similar remarks~\citep{kyledaiglegithubstaffSurveyAIWave2024}.
To prepare computing students for careers that may require them to use LLMs, many instructors are now introducing LLMs into computing courses~\citep{vadaparty_cs1-llm_2024, liu_TeachingCS50AI_2024a, zamfirescu-pereira_61ABotReport_2024}. However, this begs the question, what must students know to successfully use LLMs for programming tasks?

Recent research argues that LLMs can give students an ``illusion of competence''~\citep{prather_widening_2024}, when in fact they struggle with fundamental tasks such as writing prompts for programming~\citep{charlie}. Other researchers are developing tools and techniques to teach students how to prompt better~\citep{ma_WhatShouldWe_2024, denny_PromptProblemsNew_2024, denny_ExplainingCodePurpose_2024}.
At the same time, there are growing concerns about students' ability to understand the outputs produced by LLMs.
For example, \citet{rahe_HowProgrammingStudents_2025} observed that Introduction to Computer Science (CS1) students often repeatedly submitted LLM-generated solutions without modification, suggesting that they accepted the code without fully understanding it.
Building on this concern, we investigate whether CS1 students can understand the code that LLMs produce.

Professional workflows depend on understanding and reviewing LLM-generated code to ensure its correctness and reliability.
If students are unable to comprehend LLM-generated code well, the full potential of LLMs in programming will be undermined.
Yet, typical programming courses still rely on a ``writing-first'' approach~\citep{nelson_xie_ko_2017}: students learn to write code first, before being taught to read others' code (if at all). 
Likewise, previous work on CS student-LLM interactions~\citep{charlie, denny_PromptProblemsNew_2024, prather_widening_2024} employs automated testing to check correctness of LLM-generated code, unintentionally perpetuating a focus on whether the code passes test cases rather than whether students truly understand how it works.
Thus, we ask: \emph{are CS1 students ready to integrate LLM-generated code into real-world programming workflows?}
To answer this question we examine their ability to comprehend code from LLMs.

Our research investigates if CS1 students can effectively understand and review LLM-generated code.
Specifically, we examine their strategies for validating correctness across a broad spectrum of LLM-generated code, highlighting the cognitive and technical challenges they face when assessing code from LLMs.
Specifically, we aim to address the following research questions:

\textbf{RQ1} Can CS1 students understand LLM-generated code, and what attributes of the code determine their performance?

\textbf{RQ2} How do individual differences among students affect their code comprehension in the tasks?

\textbf{RQ3} What comprehension processes and challenges do students report when thinking about LLM-generated code?

In our lab-based user study, we use CS1 tasks and select code samples to represent a diverse set of code with ranging levels of complexity, correctness, and documentation to probe and capture potential student struggles.
Our study uncovers key barriers for CS1 students including unfamiliar coding styles, overconfidence in code accuracy, and limited programming experience; this issue arises for CS1 students in general, irrespective of personal background and experience with using LLMs.
Based on our findings, we recommend that CS1 educators adopt LLM-aware teaching strategies to account for the emerging challenges. Conversely, developers of LLM-based products should keep beginner programmers in mind while developing tools.
\footnote{Data collected from this study is available at \url{https://doi.org/10.17605/OSF.IO/4ZVJM}}
\section{Related Work}

\subsubsection*{CS Student-LLM Interactions}
This section surveys representative works on CS student-LLM interaction, examining tasks such as prompt writing, error explanation, and code assistance. In our work, we specifically focus on student evaluation of the code itself.

Prather et al. \cite{prather_widening_2024} observed that many novice programmers rely on LLMs for writing code and provide hints. They concluded that there is an ``unfortunate divide'' in the use of LLMs between students who struggled and those who did not.
In their study, although 20 out of 21 students successfully completed the assigned task with LLMs, those who struggled at various points in the study over-evaluate their problem solving ability, exhibiting an ``illusion of competence''. 
This effect arises as students who face challenges tend to offload critical problem-solving responsibilities onto LLMs, potentially hindering their development of essential programming skills.

Other recent work focused on studying CS1 students' abilities to write prompts for LLMs~\cite{charlie}.
Leveraging a carefully designed interface, researchers evaluated 120 CS1 students writing textual prompts to solve programming problems at the CS1 level across 3 institutions.
Findings revealed that beginning programmers continue to face challenges in solving programming problems, with an mean eventual success rate of only 57\%.
The study's main finding revealed that effective prompting of LLMs required a degree of prior programming experience and understanding of programming terminology.
Their observation highlighted the continued necessity of understanding computing knowledge even when using LLMs;
subsequent work \cite{lucchetti_SubstanceBeatsStyle_2024} shows that prompts without key information describing intended behavior almost always fail to generate correct code.
The authors make an important conclusion: the code LLMs are trained on is mostly written by professional programmers, and expert knowledge is required to effectively complete programming tasks using LLMs.

\citet{kazemitabaar_CodeAidEvaluatingClassroom_2024} designed \textit{CodeAid}, an LLM-based programming assistant that helps CS students with code-related tasks, such as explaining, debugging, and generating pseudo-code.
The authors deployed CodeAid in a second-year programming class with 700 students over a semester.
Through thematic analysis of usage logs, weekly surveys, and interviews with students and instructors, they identified key patterns in usage, perceptions, and LLM response quality.
The authors provided recommendations for designing AI assistants tailored to educational contexts.
We similarly use thematic analysis to identify students' experiences and challenges with our code comprehension tasks.

\subsubsection*{Code Comprehension for Human-Written Code}
While the research above exemplifies the recent focus on LLM-generated code, there is an existing body of work focused on code comprehension of programs written by humans without LLM intervention.
Prior studies have examined participants' ability to predict the output of human-written functions~\citep{Bauer_2019, Ajami_Woodbridge_Feitelson_2019, Peitek_2018}.
Various alternate measures of code comprehension has been proposed, including program summary~\citep{Bednarik_2006, McChesney_2019}, answer comprehension questions~\citep{Avidan_2017} and bug identification~\citep{Castelhano_2019}.

Unlike our study, these prior work did not specifically recruit CS1 students; instead, their participants were primarily advanced students or professionals.
A key distinction between (near-)professional and educational settings is the ability to assume a baseline level of programming knowledge.
For instance, tasks such as bug identification generally require a deeper level of understanding of programming concepts, making them less suited for beginners.
Our study, which recruited specifically beginner programmers with a strict limit on programming experience, fills a gap in the literature in the study of code comprehension for this specific demographic.

% \subsubsection*{Problem Comprehension for Students}

% \citet{wrenn_ExecutableExamplesProgramming_2019} investigated the effect of example-first problem solving strategy. 
% In an accelerated introduction to computer science course, students are required to submit test cases for coding problems.
% Test cases are verified for validity (i.e. test cases should make correct programs pass) and thoroughness (i.e. test cases should make incorrect programs fail).
% They found that this method of pedagogy improved student test cases' validity without sacrificing in increase of size and thoroughness.
% Our study also contain a problem comprehension component, yet we only use it to provide context for the code comprehension part that follows.

% we seem to be out of space -- I think this one is less relevant than the previous work, thus removing it.
\subsubsection*{Value of Buggy Code}

\citet{dibia_aligning_2023} argue that buggy code produced by an LLM can still be useful \emph{to an expert} when it is close enough to a desired solution.
They argue that an expert can often correct small mistakes faster than they can write code from scratch.
However, their work presupposes that the user can quickly identify errors in LLM generated code -- we show that CS1 students struggle to do so.

\section{Experimental Design}
\label{app:experimental-design}
The user study consists of 5 output prediction tasks for each participant.
Each task is sourced from a CS1 problem, and presents (1) a CS1-student-written prompt collected by \charlie \cite{charlie} with a sample input-output pair, and (2) an LLM-generated code that was prompted to solve the problem.
The tasks are tailored to be challenging, yet accessible to CS1 students.
This approach allows us to observe student code comprehension in a controlled environment with problems whose difficulty is tailored to their current understanding.

\subsection{Definitions and Numerical Overview}
\label{subsec:dataset-and-task-definitions}
This study relies on a structured dataset of CS1-level programming problems, paired with diverse LLM-generated code variations and corresponding test cases.
For the sake of clarity and consistency, we define key terms relevant to our experimental setup and provide a numerical overview:
\begin{itemize}
    \item \textit{Problem}: A CS1-level programming question, selected from a curated set of 48 problems~\cite{charlie}. 
    Each problem includes a prompt, 3 test cases, and an additional test case for context.
    \item \textit{Test case}: Predefined input-output pairs used to evaluate expected program behavior.
    \item \textit{(LLM-Generated) Code variation}: A function produced by Code Llama~\citet{roziere_CodeLlamaOpen} in response to a prompt solving the problem. There are 8 distinct code variations.
    \item \textit{Prompt}: A student-written textual description of the problem, collected from the study in~\citet{charlie}.
    Only prompts with a high pass@$1$ rate ~\citep{chen_valuatingLargeLanguage_2021}($\geq$ 90\%) were included.
    \item \textit{Task}: A 3-\textit{part} structured evaluation unit consisting of a problem, a corresponding code variation, and 3 test cases.
    Each participant completes 5 tasks, each based on a different problem.
    There are 40 unique tasks in total, as there are 5 problems and 8 code variations.
    \item \textit{Data point}: A single participant's response to a task, including their predictions for both expected and actual function outputs given the inputs of the 3 test cases.
    Given that each participant completes 5 tasks and we recruited 32 participants, the study generates 160 data points in total. That is, we have collected 480 ($160 \times 3$) participant-generated input-output predictions across all tasks.
\end{itemize}

\subsection{Study Dataset}
\label{subsec:study-dataset}
To ensure the study's validity and relevance, we applied constraints to curate diverse LLM-generated code and high-quality problem prompts. This section outlines the criteria and processes used to construct the \textit{study dataset}, comprising problem prompts and LLM-generated code exhibiting varied characteristics.

\subsubsection*{Choice of LLM}
For this study, we used Code Llama, to select prompts and generate code.
While multiple LLMs were considered, we opted for a single model to maintain a controlled experimental scope.
This approach helps isolate comprehension challenges specific to LLM-generated code, avoiding variability introduced by differences in model behaviors.
At the time of dataset design (June 2024), Code Llama was one of the strongest publicly available code generation models, making it a suitable choice.
Since the focus of our study is CS1 students' interactions with generated code, and not the code itself, we did not prioritize models optimized for general problem-solving or chain-of-thought reasoning.

\subsubsection*{Prompt Selection Criteria}
Given that the presented code might be buggy, ambiguous, or difficult to understand, the prompts play a pivotal role in shaping student understanding of the task. 
While prompt quality can be assessed through various dimensions~\citep{charlie}, we focused on prompts that reliably generate code passing test cases.
Specifically, a \textit{high-quality} prompt is defined as one achieving a \textit{pass@$1$} rate of 90\% or higher when tested on Code Llama.
This criterion minimizes confounding factors related to unclear or inadequate prompts.
Each problem in the study required at least one high-quality prompt, from which a representative was selected for the task.

\subsubsection*{LLM-Generated Code Selection Criteria}

There are many ways to classify code. 
For this study, we hypothesized three characteristic classes based on our assumptions regarding students' potential areas of struggle. 
The characteristic classes are:
\begin{itemize}
    \item \textit{Correctness}: Whether the function correctly solves the given problem, determined empirically by running test cases.
    \item \textit{Comment presence}: Whether the function text includes comments generated by the LLM.
    \item \textit{Complexity}: Whether the function contains iteration such as loops, list comprehension, or recursion.
\end{itemize}
To ensure diversity, we selected problems that have LLM code generations representing all possible combinations of these characteristic classes, resulting in 8 different function variations. 
See Figure \ref{fig:reversewords1} for an example of correct, non-complex implementation of the \texttt{reverseWords} problem with comments and Figure \ref{fig:reversewords2} for an incorrect, complex implementation with comments for the same problem.
This approach provides participants with exposure to a comprehensive range of scenarios.

% \begin{figure}[h]
% \centering
% \begin{minipage}{0.45\textwidth}
% \centering
% \begin{lstlisting}
% def reverseWords(words):
%   # reverse each word in the list
%   for i in range(len(words)):
%     words[i] = words[i][::-1]

%   # sort the list
%   words.sort()

%   return words
% \end{lstlisting}
% \caption{A correct implementation of \texttt{reverseWords} that reverses each word individually and sorts the list after. This function is categorized as comments and non-complex.}
% \label{fig:reversewords1}
% \end{minipage}%
% \hspace{0.05\textwidth}%
% \begin{minipage}{0.45\textwidth}
% \centering
% \begin{lstlisting}
% def reverseWords(words):
%   # reverse the words
%   words = [word[::-1] for word in words]
%   # sort the words
%   words.sort()
%   # reverse the words again
%   words = [word[::-1] for word in words]
%   return words
% \end{lstlisting}
% \caption{An incorrect implementation of \texttt{reverseWords} that did an extra reverse after sorting. This function implementation is categorized as with comments and complex (due to presence of list comprehension).}
% \label{fig:reversewords2}
% \end{minipage}
% \end{figure}

\begin{figure}[t]
\centering
\begin{subfigure}[t]{0.48\textwidth}
\centering
\begin{lstlisting}
def reverseWords(words):
  # reverse each word in the list
  for i in range(len(words)):
    words[i] = words[i][::-1]
  # sort the list
  words.sort()
  return words
\end{lstlisting}
\caption{A correct implementation of \texttt{reverseWords} that reverses each word and sorts the list. Categorized as non-complex with comments.}
\label{fig:reversewords1}
\end{subfigure}
\hfill
\begin{subfigure}[t]{0.48\textwidth}
\centering
\begin{lstlisting}
def reverseWords(words):
  # reverse the words
  words = [word[::-1] for word in words]
  # sort the words
  words.sort()
  # reverse the words again
  words = [word[::-1] for word in words]
  return words
\end{lstlisting}
\caption{An incorrect version that unnecessarily reverses words again after sorting. Categorized as complex (due to list comprehension) and with comments.}
\label{fig:reversewords2}
\end{subfigure}
\caption{Two implementations of \texttt{reverseWords}.}
\end{figure}

\subsubsection*{Dataset Construction}
To construct the dataset, we drew from the set of 48 CS1-level problems designed for the user study of~\citet{charlie} and a subset ($n=947$) of prompts collected for those problems.
Prompts were written by participants with only one CS course completed at the time of the study.
We then prompted Code Llama with these prompts and obtained 42,702 unique Python functions.
After filtering for completeness and prompt quality, we identified 9 problems that met our criteria for diverse code examples and high-quality prompts.
From these, we selected five problems (\texttt{create\_list}, \texttt{combine}, \texttt{altText}, \texttt{multisplit}, and \texttt{reverseWords}) to ensure coverage of varied data structures and types encountered in CS1. Each problem has eight characteristic-based code variations, yielding 40 unique problem-code pairings.
While an attentive reader might note that the code may not directly stem from the prompt, this is by design; the focus of the study is on the code, with prompts acting merely as a contextual framing for the problem to participants.

\subsection{Task Design}

Using this dataset, we designed structured tasks to assess beginner programmers' comprehension of LLM-generated code.
Each task consists of three \textit{parts}: \textit{prompt comprehension}, \textit{code comprehension}, and \textit{reflection} (referred to as \textit{Part A}, \textit{Part B}, and \textit{Part C}, respectively).

Participants begin with \textbf{prompt comprehension} (Part A), where they read a problem description with example inputs and outputs.
To confirm their understanding of the problem, they predict the expected output for 3 given inputs.
Next, in \textbf{code comprehension} (Part B), participants analyze an LLM-generated function intended to solve the problem. They predict its output for the same inputs as in Part A, noting that the function’s output may differ from the expected results.
Finally, in \textbf{reflection} (Part C), participants compare their predictions with the ground truth: the expected output from Part A and the actual output from Part B. They then answer questions about their experience.

\begin{figure}
    \centering
    \includegraphics[width=\linewidth]{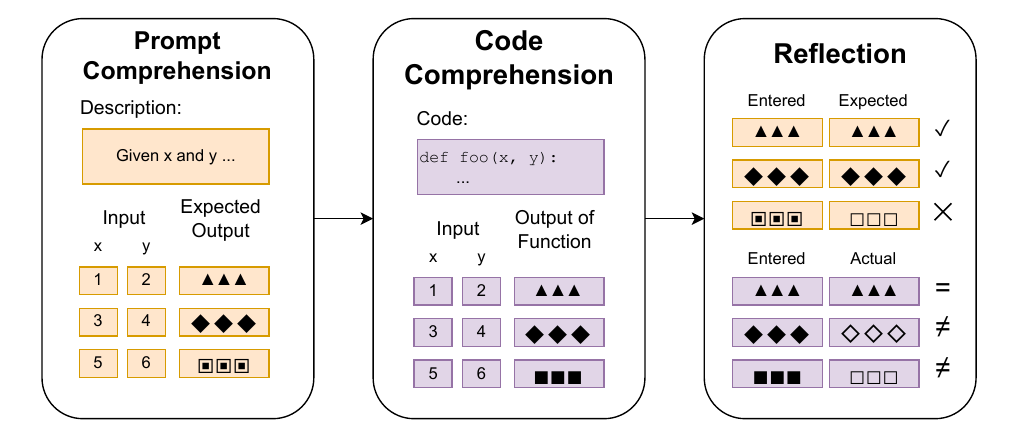}
    \caption{Procedure of a task. \textit{Prompt comprehension} presents a description collected from \charlie \cite{charlie} and three inputs. A participant enters their expected output for the described function given the description. \textit{Code comprehension} presents code generated for the problem by an LLM and the same set of inputs, but asks for the output of the code. \textit{Reflection} presents the correct answer for both parts and prompts the user to review their answer. For \textit{prompt comprehension} responses, a response is correct (\checked) if it matches the expected output or wrong ($\times$) otherwise. For \textit{code comprehension} responses, a matching response to the actual output is said to be consistent ($=$); a non-matching response is inconsistent ($\not=$) to the actual output of the code.}
    \label{fig:task-structure}
\end{figure}

Figure \ref{fig:task-structure} illustrates an example of this task, including the steps and interactions involved.
In the figure, a participant entered their predicted expected outputs and outputs of the function in prompt comprehension and code comprehension respectively. When reviewing their result, they found that their first two predicted outputs are \textit{correct}, but their third predicted output is \textit{incorrect} because it didn't match the canonical output. They also found that while their first predicted output for code is \textit{consistent} with the actual output, the second and third outputs is \textit{not consistent} with the ground truth.

We believe this set of interactions to be a sound approach for gauging students' code comprehension.
Presenting students with a prompt and an input-output pair first ensures that they contextualize the code aiming to solve a specific problem, mirroring how they naturally interact with LLMs when seeking code solutions.
Without this step, students might interpret the code arbitrarily rather than evaluating it in relation to an expected behavior.
Requiring students to explicitly enter output predictions, rather than selecting from multiple choices, prevents guessing biases and encourages mental code execution.
Additionally, compared to free-response explanations of code behavior, output prediction is a more direct and objective measure, eliminating interpretation variations and ensuring efficiency in evaluation.
By structuring the task this way, we ensure that comprehension is tested in a way that is both rigorous and aligned with real-world reasoning about code, where developers frequently predict outputs to verify their understanding of a function’s execution.

The accuracy of participants' predictions serves as an objective measure of their understanding of both the problem description and the generated code. 
To supplement this quantitative assessment, we collect additional insights by asking participants to rate their confidence, along with their predictions for both prompt and code comprehension.
Furthermore, they are invited to reflect on their experiences and their understanding across both parts, after reviewing their results against the ground truth.
More specifically, participants are asked to indicate if the generated code is what they themselves would write, as done in \charlie \cite{charlie}.
Following it is a free-response section, inviting them to describe their experience of the task.
The exact wording of the free-response section can be seen in Figure \ref{fig:study-interface-partc}.

This open-ended format allows participant to articulate specific challenges they faced, such as difficulties in understanding the prompt or part of the code.
In summary, this combination of quantitative and qualitative data provides a comprehensive view on the challenges involved in understanding LLM-generated code for beginning programmers.

\subsection{Study Interface}

\begin{figure*}[ht]
    \centering
    % Subfigure 2
    \begin{subfigure}[b]{0.48\textwidth}
        \centering
        \includegraphics[height=8.5cm]{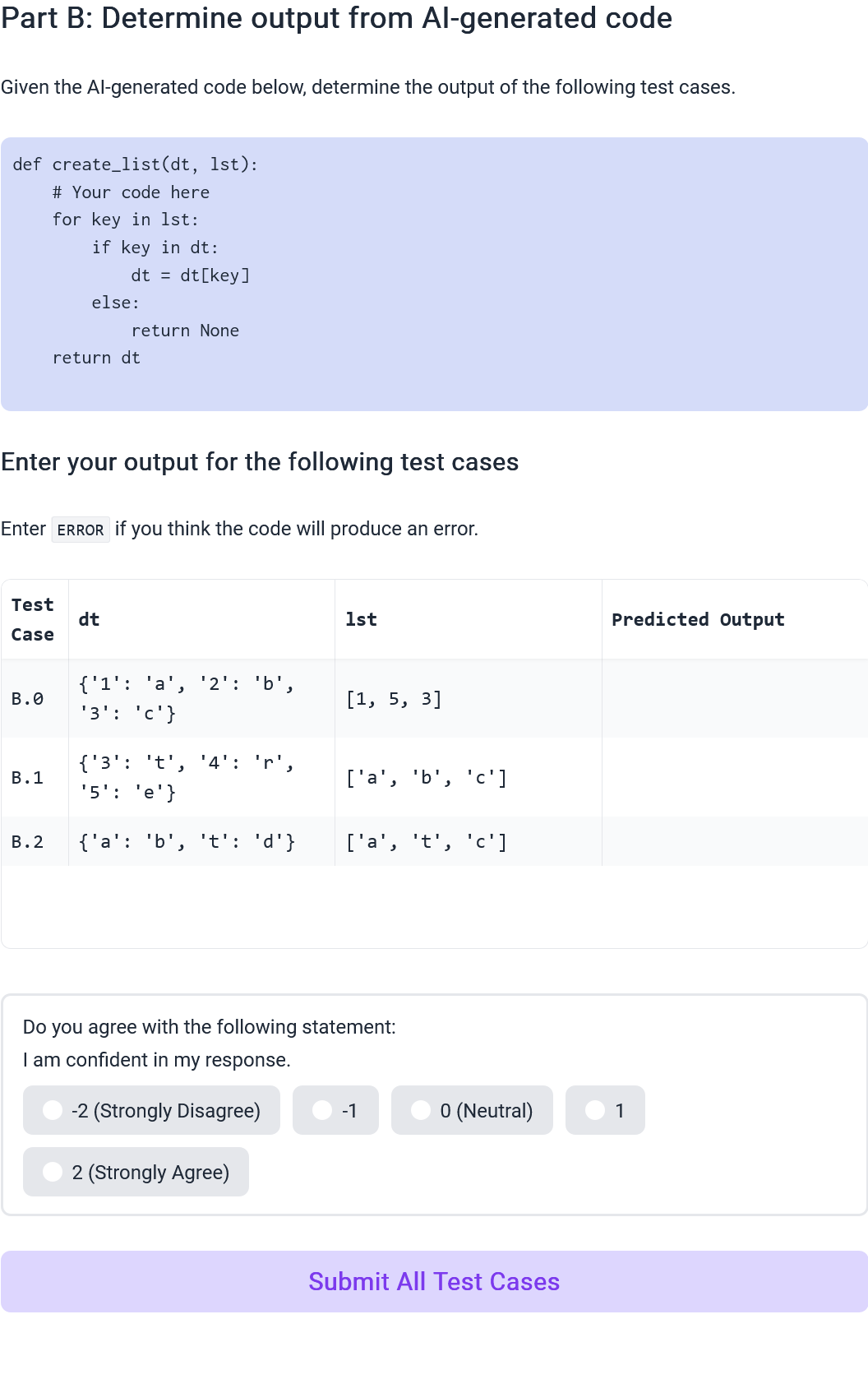} % Replace with your image file
        \caption{Part B of the example task shown to participants.
The interface shows a prompt in the top section and LLM-generated code in the code section.
Participants read the inputs in the table and write their predicted outputs in the last column.
Each test case is labeled (e.g., “B.1”, “B.2”) for reference.
Afterward, they rate their confidence using a 5-point Likert scale.
Part A and Part B share the same structure.
        %their agreement to the statement ``I am confident in my response'' in a 5-point Likert scale ranging from -2 (``Not Confident'') to 2 (``Confident'').
        %Since both parts are similar in structure, Part B is presented here as an illustration.}
        }
        \label{fig:study-interface-partb}
    \end{subfigure}
    \hfill % Adds space between subfigures
    % Subfigure 3
    \begin{subfigure}[b]{0.48\textwidth}
        \centering
        \includegraphics[height=8.5cm]{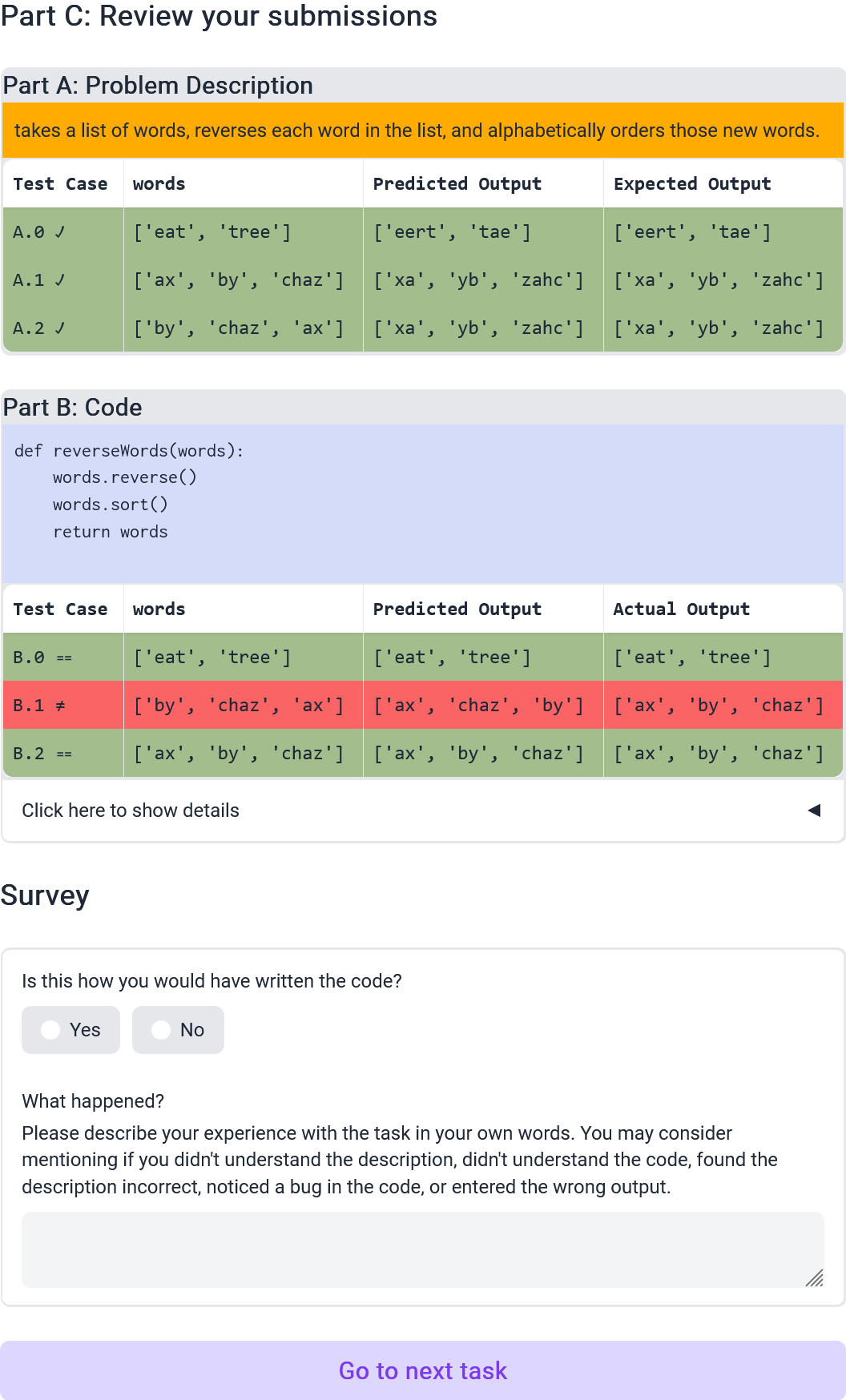} % Replace with your image file
        \caption{Participants are shown the results. Accurate and inaccurate predicted results are highlighted in different colors. 
        Participants are then prompted to answer the survey questions shown at the bottom of the page.}
        \label{fig:study-interface-partc}
    \end{subfigure}
    \caption{The Study Interface.}
    \label{fig:study-interface-1}
\end{figure*}

The study interface is designed as a web application.
We provide two interface screenshots in Figure \ref{fig:study-interface-1}. Figure \ref{fig:study-interface-partb} illustrates the interface for prompt comprehension and code comprehension (with Part B used as an example), and Figure \ref{fig:study-interface-partc} illustrates the interface for reflection.

In prompt comprehension, the participants see the prompt being presented, followed by a table displaying test case inputs. 
Participants then are prompted to enter the expected output of described function in the rightmost column of the table.
To avoid potential typos, they are required to enter syntactically valid Python values before proceeding.
Participants are asked to rate their confidence for their responses using a 5-point Likert scale, indicated via a radio button below the table, before proceeding to the next part.

The interface for code comprehension is structurally similar to prompt comprehension above. The test case inputs from prompt comprehension are used but shuffled and relabeled (e.g. ``B.1'', ``B.2'' etc.) to mitigate ordering effects.
Participants are instructed to enter \texttt{ERROR} if they believe the code raises an exception for a given input.

In reflection, the interface displays the prompt, the generated code, the participant's predicted outputs, and the expected output or actual output, for Part A and Part B respectively.
A ``Click here to show details'' expendable section provides additional explanations, showing a message regarding the correctness of the input and additionally, if applicable, the exception raised by the Python interpreter.
Participants are then asked to indicate if the generated code is what they would write to solve the given problem using a pair of radio button with yes and no as choices.
Following this question, they are required to enter their reflection on the experiences in the task in the text box below.

To ensure familiarity with the study interface, participants complete a tutorial task before proceeding to the main study tasks.
The tutorial task is designed to be less challenging than the main tasks, and we provide additional guiding text to assist participants for understanding the workflow of the study interface.

\subsection{Post-Study Survey}

After the main study, the participants are then presented with a post-study survey. \footnote{Study materials available at \url{https://doi.org/10.17605/OSF.IO/4ZVJM}}
This part takes direct inspiration from the interview and post-study components of the studies in \charlie \cite{charlie} and \citet{feldman_non-expert_2024}.
The goal of the survey is to assist in answering RQ2 and RQ3, which requires collecting qualitative data. 
%While our study use the term ``LLM'' throughout, we recognized that some students might be unfamiliar with this terminology.
In our post-study survey, we used the more approachable term ``Generative AI'' (or ``GenAI'') to reduce potential confusion with the term ``LLM''.

The first section of the survey gathers data on participants' prior usage of LLMs for code.
The participant first indicates whether or not they have ever used a LLM for coding purposes.
If they answer yes, they will then report their usage frequency and typical tasks (e.g. generating code, debugging code, etc.).
Otherwise, participants are prompted to explain the reasoning behind their non-usage in a free-text response.

The second section of the survey asked about the study experience, including five questions about their approach to checking code correctness, helpfulness of the provided examples, clarity of the prompt presented, any challenges they encountered understanding the problems, and ease of reading the code.
Participants could also share additional comments.

The third section sought to obtain participant's perceptions of LLMs for code and their broader perspectives on AI. We adapted these questions from~\citet{charlie}, modernized them to describe LLM-powered programming tools in general in the wake of their availability at the time this experiment took place (November to December 2024).

Finally, demographic and background questions were placed at the end to mitigate stereotype threat~\citep{noauthor_ncwit_nodate}, following standard practices.
% We aimed to alleviate stereotype threats~\citep{noauthor_ncwit_nodate} by putting questions regarding demographic and various background information at the last section of the survey, following standard practices. 

\subsection{Logistics}

%Due to recruitment constraints,
% we recruited 32 participants for the main study, all of whom met the eligibility criteria: being at least 18 years old, a current student at \neu{} and having completed CS1, with no additional completed CS courses.
% However, they could be concurrently enrolled in another CS course at the time of the study.

We recruited 32 participants for the main study, all of whom met the eligibility criteria: being at least 18 years old, a current Northeastern University student, and having completed DS2000—an introductory Data Science course equivalent to CS1. DS2000 is taught in Python, the same language used in our study. Participants had not completed any additional CS or DS courses, though they could have been concurrently enrolled in another course at the time of the study.

%As described in Section \ref{subsec:dataset-and-task-definitions}, each participant completed 5 tasks, ensuring coverage of all 40 unique problem-code combinations. This resulted in 160 total data points and 480 input-output predictions.

Prior the main study, we conducted a pilot study of 20 participants who had completed at least 2 CS courses.
The pilot helped refine the study materials, particularly the survey design. 
Notably, some survey questions that were related to each other shared the same input box.
After examining the survey data, we found that a number of participants answered only one part of these composite questions, without answering the other.
We decided to break these questions into separate survey questions in the main study.

The studies were run in open lab setting, with several attendees per session.
For the sake of reducing participant stress, they were told--both verbally and in the consent form--that the study would not become part of their educational record. They were also reminded that the goal of the study was not to evaluate their programming skills, similar to \charlie \cite{charlie}.
Our study was deemed exempt by the Institutional Review Board of \theneu{}. 
Participants in both the pilot and the main study were compensated a \$40 Amazon gift card each for their efforts.

\section{Analysis}

To evaluate participant performance, we used two measures: (1) the \textit{per-part} pass rate, based on 160 data points, where a part in a task is considered passing only if all three output predictions for the test cases are correct, and (2) the \textit{per-test-case} pass rate, based on 480 individual test cases.
In our analyses, we considered the pass rates of prompt comprehension and code comprehension parts separately.

For RQ1, we analyzed the coding task outcomes conditional on the success or failure of the description task, using per-part pass rates to mitigate dependencies between the tasks. For the quantitative analyses for RQ2, we relied on the more sensitive per-test-case pass rates to capture individual performance.
For RQ3, we performed thematic analysis for free-response survey answers and participant reflections, following the approach of \charlie \cite{charlie}. The first two authors independently coded responses and resolved discrepancies through discussion. 
Participant quotes presented in this paper have been pseudonymized and lightly edited for grammar.
Further details are available in the appendix.

Statistical analyses was conducted with $\alpha = 0.05$, employing chi-squared tests for independence, $z$-tests for proportion comparisons, $t$-tests for two-group mean comparisons, and ANOVA for multi-group comparisons, followed by Tukey's HSD for post-hoc analysis.

\section{Results}
\label{sec:results}

\subsection{RQ1: Student Comprehension and Code Attributes}

\begin{figure*}[ht]
    \centering
    \begin{subfigure}[b]{0.2\textwidth}
        \centering
        \includegraphics[width=\textwidth]{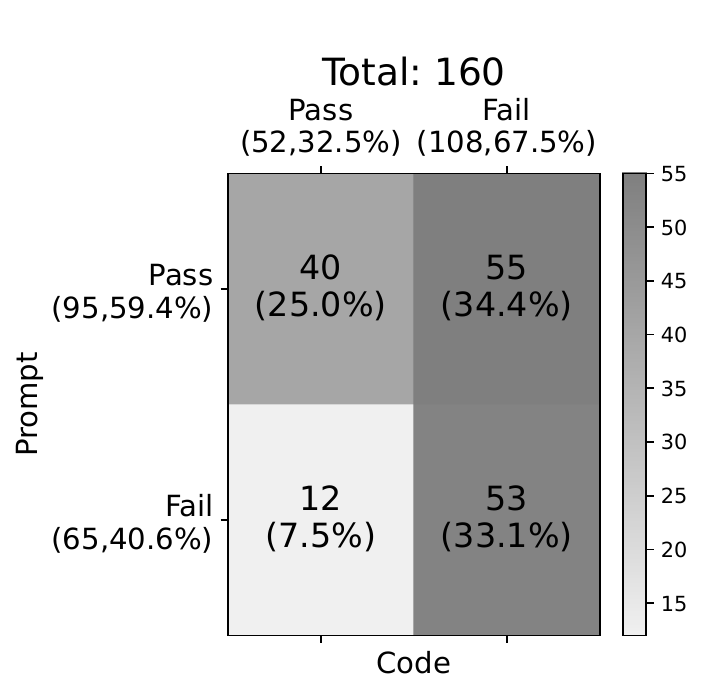}
        \caption{Success rates among all task instances.}
        \label{fig:rq1-per-task-all}
    \end{subfigure}
    \hfill % Adds space between subfigures
    \begin{subfigure}[b]{0.2\textwidth}
        \centering
        \includegraphics[width=\textwidth]{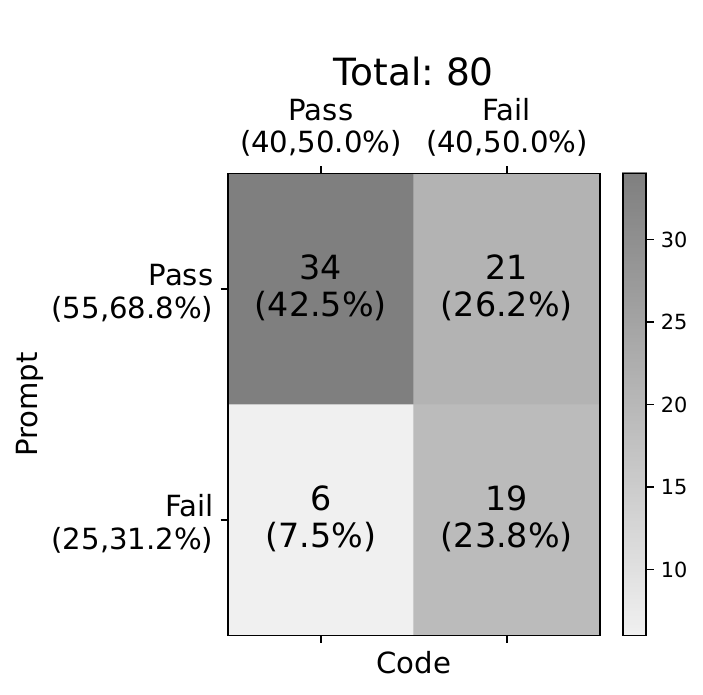}
        \caption{Success rates among task instances with correct code.}
        \label{fig:rq1-per-task-correct}
    \end{subfigure}
    \hfill % Adds space between subfigures
    \begin{subfigure}[b]{0.2\textwidth}
        \centering
        \includegraphics[width=\textwidth]{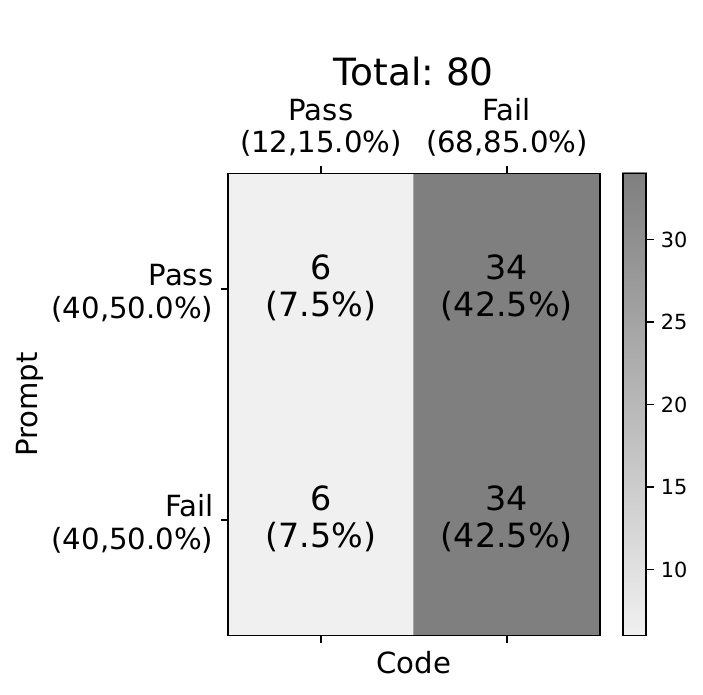}
        \caption{Success rates among task instances with wrong code.}
        \label{fig:rq1-per-task-wrong}
    \end{subfigure}
    \caption{Counts and percentages of per-part success rates for prompt comprehension (Prompt) and code comprehension (Code) under different code correctness conditions: (a) all task instances, (b) tasks with correct code, and (c) tasks with incorrect code.}
    \label{fig:rq1-per-task}
\end{figure*}

\subsubsection*{Participants are more likely to succeed at prompt comprehension than code comprehension}

% Arjun: Mention: 59% of the time, students correctly predict output on a given input when they are only reading a prompt (without code). 
% However, only 32% of the time they correctly predict output when given LLM-generated code.
We summarize our per-part results in Figure \ref{fig:rq1-per-task}. 
Participants demonstrated moderate success in prompt comprehension.
Across 160 total tasks completed, participants successfully completed 95 ($59.3$\%) of them.
In contrast, participants encountered greater difficulty with the code comprehension task, achieving success in only 52 tasks (32.5\%).
Interestingly, in 55 tasks (34.4\%), participants succeeded in the description component but failed in the corresponding code component.

\subsubsection*{Participants are more likely to fail at code comprehension when the code has bugs}

Among the 160 total tasks completed, half of the tasks had correct LLM-generated code, meaning that the code solved the problem as described in prompt comprehension (Figure \ref{fig:rq1-per-task-correct}). The other half had faulty code (Figure \ref{fig:rq1-per-task-wrong}). We performed a two sample $t$-test on the correct and incorrect group and found that the correct group has significantly higher success rate than the incorrect group, meaning that students more consistently produced correct outputs given correct code ($t = 6.219$, $p < 0.0001$).

%\subsubsection*{Lower code comprehension success rates following failed prompt parts}
\subsubsection*{Participants disproportionately fail at code comprehension, despite success at prompt comprehension}
Prompt and code comprehension success rates are statistically dependent, as shown by a per-task chi-squared test ($\chi^2 = 8.79$, $p=0.003$). 
Participants who fail the prompt comprehension task are highly unlikely to succeed in code comprehension (pass probability: $0.18$, failure probability: $0.82$; left-tailed $z$-test: $z = -7.192$, $p < 0.0001$).
Surprisingly, even when participants understand the prompt, their failure rate in code comprehension ($0.58$) exceeds their success rate ($0.42$; right-tailed $z$-test: $z = 2.176$, $p = 0.015$). This highlights a significant struggle with code comprehension, even after prompt understanding.

\subsubsection*{Unclear relationship between participant success and the choice of problem, code complexity, or presence of comments}

% A one-way ANOVA test was conducted to examine whether the problem students were given had a significant influence on task success.
% The test did not provide sufficient evidence to conclude that success rates were affected by the problem, code complexity, or presence of comments.
% Additionally, two-sample $t$-tests were performed to evaluate the effect of comment presence and complexity on participant success.
% Neither test yielded statistically significant results, suggesting that any potential effect may be too small to detect given our sample size and methodology.
% Future work could explore these relationships further with a larger sample or alternative measures of comprehension.

A one-way ANOVA and two-sample $t$-tests found no significant effect of problem choice, code complexity, or comment presence on task success. The results suggest any potential effect may be too small to detect with the given sample size and methodology. Future work could investigate these factors with larger samples or alternative comprehension measures.

\subsection{RQ2: Limited Individual Differences}

\subsubsection*{Non-native English speakers do worse at prompt comprehension}
We found a statistically significant difference between the language spoken in childhood and the success rate of prompt comprehension (ANOVA, $f = 8.5640$, $p = 0.0012$).
Participants were asked about the languages spoken in their childhood. We found that participants raised in other languages than English have significantly worse prompt comprehension performance (success rate 0.68; Tukey's HSD, $p \leq 0.002$) than participants who spoke English in their childhood ($0.82$ for multilingual with English, $0.88$ for English only).
Interestingly, the same conclusion can not be made for the performance on their code comprehension part.

\subsubsection*{Unclear relationship between participant success, self-reported experience with Generative AI, and other demographic factors}

Our analysis found no statistically significant correlation between LLM usage frequency and task success\footnote{See appendix for detailed results.}, meaning its impact remains inconclusive.\footnote{This may be influenced by course policies discouraging LLM use.}

Similarly, no significant performance differences were observed across demographic factors (e.g., gender, race, programming experience). However, potential effects may exist, but were undetected due to sample size and distribution.

% Our analysis did not provide sufficient evidence to conclude that the frequency of using LLMs had a significant impact on success rates for any task (Table \ref{tab:demographic}).
% This suggests that frequent use of LLMs for programming does not necessarily translate to an improved ability to evaluate LLM-generated code among CS1 students.\footnote{The students' computing courses discourage or forbid them from using LLMs, which may influence this result.} 

% Additionally, we examined various demographic factors (e.g., gender, race, programming experience) in relation to task performance. No statistically significant differences were found between groups (Table \ref{tab:demographic}).
% However, it is possible that effects exist but were not detected given the sample size and distribution.

% \subsubsection*{Participant success is not impacted by their self-reported experience with Generative AI and other demographic factors.}
% Surprisingly, we found that the frequency of using LLMs has no statistically significant impact on their success rates on any task (Table \ref{tab:demographic}).
% Our results indicate that frequent use of LLMs for programming does not appear to enhance CS1 students' ability to evaluate LLM-generated code.\footnote{The students' computing courses discourage or forbid them from using LLMs.}
% We also inspected a number of other demographic factors (e.g. gender, race, programming experience, etc.) with performance in the task and found no statistically significant difference between all groups (Table \ref{tab:demographic}).

\subsection{RQ3: Student Processes and Challenges}

\subsubsection*{Familiarity with LLMs for Code}
The post-study survey revealed that 24 (75\%) participants already use LLMs for coding tasks. 
Among these respondents, 15 (46.9\%) modify code, 22 (68.6\%) debug code, 20 (62.5\%) use LLMs to explain code, and 10 (31.25\%) students use LLMs to generate code.
Notably, all but one of the above participants use LLMs for more than one code-related purpose.
For the 8 (25\%) respondents who do not use LLMs for coding tasks, reasons included concerns about plagiarism ($n = 2$, 6.25\%), AI over-reliance ($n = 4$, 12.5\%), distrust in correctness of AI responses ($n = 1$), and the inability of AI to fully understand users ($n = 1$).

\subsubsection*{How do students check the correctness of LLM-generated code?}
Our post-study survey asked participants about their approaches to checking code correctness during the study.
Many participants ($n=19$, 59.4\%) rely on code tracing as the main strategy.
Some participants mentioned using either the provided example ($n = 5$, 15.6\%) or examples of their own ($n = 3$, 9.4\%) while tracing code.

\subsubsection*{Why is code from LLMs hard to understand?}
We solicited feedback from participants about their experiences and challenges during the code comprehension task, both throughout the task itself and in the post-study survey.

The largest theme category reported is attributed to a lack of understanding the code due to 3 factors.
18 participants (56.3\%) reported that the code presented to them contained Python features (e.g. list comprehensions and list slicing) they are not familiar with.
For instance, \textsc{livelyNomad} commented when facing list slicing index in Python: ``\textit{I did not know what \texttt{word[::-1]} did as I have never seen two colons used like that before.}''
13 participants (40.6\%) pointed out instances of not understanding library functions in the generated code. \textsc{boldGlacier} pointed out: ``\textit{I wasn't sure what \text{[\texttt{list}]}\texttt{.join()} does}.''
7 participants (21.9\%) reported having trouble understanding certain fundamental programming concepts.  
\textsc{livelyLion} said after reviewing code that recursively calls itself: ``\textit{I noticed that the function is trying to recall itself, which might be right, but not how I write my code, and I think is wrong, so it caused my answer to be all errors.}''

16 participants (50.0\%) reported encountering situations where they perceived a code snippet as wrong, but were unable to predict its outputs.
\textsc{fearlessStar} noticed this issue, yet misattributed the inconsistency to the sample input-output pair: ``\textit{the example was incorrect. [...] It said that every other letter alternated to be upper/lower [case] but there were two instances in new words where the letters were the same case [...] .}''
Of these 16, 10 participants (31.3\%) were able to point out exactly where the inconsistencies between the code and the prompt lay, whereas the remaining 6 (18.8\%) were able to indicate that there was a bug, yet were unable to relate the bug to the intended behavior provided by the prompt. 

14 participants (43.8\%) commented on the style of the code they encountered. \textsc{fearlessStar} noted their unfamiliarity with the style of a function utilizing a list comprehension: ``\textit{The only thing I would have done differently is put the \texttt{for i in range} before the function because that's what we were taught in class. I never did it the way the AI generated.}''
\textsc{gracefulKnight} observed that the code style initially hindered their understanding, even though they eventually understood it: ``\textit{The code was written in a single line that made it hard to understand what exactly was being returned, but it was readable after looking further into it.}'' 
\textsc{livelyNomad} made the same observation, yet noted a more positive remark: ``\textit{I thought this code was really interesting as most of the code presented I would have assumed would not work at all. I definitely would have written my code differently as would probably every single other programmer at my level.}''

Participants indicated that either a lack of comments ($n = 4$, 12.5\%) or an excess of comments ($n = 5$, 15.6\%) hindered their code comprehension. \textsc{shinyOak} encountered code with commented-out lines containing unused code. They say: ``\textit{Having all the commented out lines is bad to look at.}''

Participants' understanding of the problem description could mislead them when interpreting LLM-generated code.
For 12 task instances (out of 160, 7.5\%), participants pointed out that problem descriptions were difficult to follow (despite their effectiveness as LLM prompts). 
5 participants (15.6\%) brought up similar issues in the post-study survey.
For instance, \textsc{livelyForest} remarked in the post-study survey that they ``\textit{had some problems understanding what the problems wanted [them] to do}'' which eventually ``\textit{may have skewed [their] results for the generated code}''.

Only 3 participants (9.3\%) explicitly stated that when they made a mistake predicting code outputs, it was because they assumed the code would work as described in the problem comprehension task. 
However, upon examining participant-entered outputs, we found that 29 participants (90.6\%) entered the same output in prompt comprehension and code comprehension parts at least once, despite the fact that the entered output was inconsistent with the actual execution result of the code.
This discrepancy suggests that participants frequently assumed that the code was correct.

\subsubsection*{The effect of task setup}

In the post-study survey, we also asked students about if the descriptions (i.e. prompts) were clear and if the examples were useful.
All participants indicated that examples were useful, but only 26 participants (81.3\%) found that the prompts were clear in general. 
8 (25\%) participants pointed out that they perceived inaccuracies in the prompts. 

\section{Discussion}

We begin by noting that while most students were able to understand the problem itself from a combination of prompts and examples, they report struggling to understand  the LLM-generated code that attempts to solve the problem. Notably, this manifests in students disproportionately misunderstanding LLM-generated code compared to natural language prompts, as shown in our quantitative analysis. 

Our post-study survey data showed that code misunderstandings are consistent across CS1 students, regardless of demographics factors or prior AI experience. 
We identify three main factors that have contributed to this phenomenon: unfamiliar code style, incorrect assumption of correctness, and lack of expertise.
However, non-native English speakers demonstrated an additional disadvantage in prompt comprehension, suggesting an important equity concern that aligns with existing studies on accessibility barriers \citep{molina_LeveragingLLMTutoring_2024, babe_StudentEvalBenchmarkStudentWritten_2023}.

In our survey responses, almost half the participants explicitly state that the LLM-generated code was in an unfamiliar style or syntax, or used approaches or built-in functions that they did not know. Reading the code ourselves, we determine that in all of these cases the LLM typically produces succinct idiomatic Python solutions.  However, students with just one semester of Python are not typically taught to write or read ``Pythonic'' code, for general structures (e.g. loops) are likely taught in favor of Python-specific expressions (e.g. list comprehensions).

We also find that students struggle to predict the output of buggy LLM-generated code.
We hypothesize that this occurs because participants exhibit automation bias~\citep{skita_accountability_2000,goddard_automation_2012, gadala_automation_2017, de-arteaga_case_2020}, believing that the code is correct, even though we explicitly remind them that the code may have bugs and that the point of the study is to discover them.
In other words, participants often fail to predict the behavior of the code in front of them, and instead reason about the behavior of the intended function.
This phenomenon could contribute to students' ``illusion of competence''~\citep{prather_widening_2024} and highlights the need to teach code comprehension and debugging strategies.
% the code comprehension task with
% approach the code task with an existing assumption about the correctness of the code, and are thus writing outputs according to the specifications of the description task, rather than according to the code provided. This presumption about correctness of code is a likely contributor to the disparity in success rates we see in the analysis of RQ1. 

% the correctness of the LLM generated code is a significant factor to the success rate of code comprehension, and that students are on average achieving higher success rates in tasks where the code is correct. Though not conclusive, this result leads us to hypothesize that participants approach the code task with an existing assumption about the correctness of the code, and are thus writing outputs according to the specifications of the description task, rather than according to the code provided. This presumption about correctness of code is a likely contributor to the disparity in success rates we see in the analysis of RQ1. 

We also find that participants are much more likely to fail at code comprehension when they also fail at prompt comprehension.
This fact suggests that first understanding the problem is critical to understanding code and highlights a potential hazard with using LLMs: they can produce incorrect code even when the user does not fully understand the problem.
The ability to generate the code quickly and effortlessly with LLMs exacerbates this issue, making it even more challenging to catch and correct mistakes.

%Our study revealed a dependence between the success rate of prompt comprehension and code comprehension tasks, and showed that when students fail the prompt comprehension part, their success rate for code comprehension is significantly lower. 
%These findings suggest that a fundamental understanding of the problem is critical for comprehending code that addresses the problem, and we hypothesize that a lack of expertise from beginner students is also a likely contributor to the misunderstanding we see from quantitative results. This point also raises concern on the use of LLM coding tools early in a students' career. Using LLMs to generate and write code without fundamental understanding of programming may mean that students are both unable to verify the correctness of the code generated, and unable to reap the benefits of deepened understanding and learning.

Finally, participants report that reading input/output examples was often more helpful than reading natural language prompts of the problem. Thus we believe that the practice of learning to write and reason about input/output examples should continue to be actively taught, as work prior to the advent of LLMs has consistently advocated~\citep{felleisen2018design,edwards_ImprovingStudentPerformance_2003, carrington_TeachingSoftwareTesting_1997}.

\section{Threats To Validity}

The choice of LLM may influence generated code quality and, consequently, student responses.
Future work should explore this task with additional LLMs to assess how different models affect code comprehension, providing a broader perspective on their impact in educational settings.
Additionally, as LLMs rapidly evolve, they may better accommodate to the needs of beginner programmers, potentially addressing challenges identified in this study~\citep{liu_TeachingCS50AI_2024a, feng_CourseAssistPedagogicallyAppropriate_2024}.

Our study used the success rates--specifically, entering correct output values for Python code--as a proxy for code comprehension.
Nonetheless, this measure could be flawed, for the ability to predict code output may not perfectly correlate to code understanding.
For instance, participants occasionally recognized that the code did not solve the problem, yet were still unable to provide the correct output.
To address this limitation, we collected qualitative data and incorporated it into our study, gaining a partial understanding of participants' thought processes and challenges in their perspective.

A threat to external validity exists:
our data collection is only conducted at one institution with one population.
Future work should consider replication across multiple institutions with diverse student populations.

\section{Conclusion}
\balance
Our study has uncovered a trend among CS1 students: they struggle more to understand LLM-generated code than natural language prompts describing intended behavior.
Specifically, they frequently misunderstand LLM-generated code due to their unfamiliarity with code style, overestimation of correctness and limited programming proficiency.

Qualitative analysis revealed that code comprehension, especially of code produced by AI, poses unexpected challenges on multiple levels.
These findings underscore the complexity of integrating Code LLMs into CS1 curricula.

Our investigation reaffirms observations from many other studies on LLMs~\citep{tian_CodeHaluInvestigatingCode_2025,tyen_LLMsCannotFind_2024,wang_AssessingFactualReliability_2024,li_LeveragingLargeLanguage_2024,schroeder_CanYouTrust_2024} across various research communities:
LLM users need to meticulously scrutinize model output. 
This emphasizes the importance of developing critical evaluation skills in CS1 students, enabling them to assess AI-generated code without relying solely on external tools. Future work should focus on improving student's ability to interpret and evaluate AI-generated code.

\begin{acks}
This work is supported by the National Science Foundation (SES-2326173, SES-2326174, and SES-2326175).
We also would like to thank many people who extended help in testing the early prototype of the study platform. 
\end{acks}

% this is only cited in the appendix/supplementary materials, but doing this just to make sure that it also appears in the list of main doc
\nocite{hart_development_1988}

\bibliographystyle{ACM-Reference-Format}
\bibliography{biblio}


\begin{thebibliography}{42}

%%% the Web address.

\ifx \showCODEN    \undefined \def \showCODEN     #1{\unskip}     \fi
\ifx \showDOI      \undefined \def \showDOI       #1{#1}\fi
\ifx \showISBNx    \undefined \def \showISBNx     #1{\unskip}     \fi
\ifx \showISBNxiii \undefined \def \showISBNxiii  #1{\unskip}     \fi
\ifx \showISSN     \undefined \def \showISSN      #1{\unskip}     \fi
\ifx \showLCCN     \undefined \def \showLCCN      #1{\unskip}     \fi
\ifx \shownote     \undefined \def \shownote      #1{#1}          \fi
\ifx \showarticletitle \undefined \def \showarticletitle #1{#1}   \fi
\ifx \showURL      \undefined \def \showURL       {\relax}        \fi
% invisible to TeX
\providecommand\bibfield[2]{#2}
\providecommand\bibinfo[2]{#2}
\providecommand\natexlab[1]{#1}
\providecommand\showeprint[2][]{arXiv:#2}

\bibitem[Q3E(2024)]%
        {Q3EarningsCall2024}
 \bibinfo{year}{2024}\natexlab{}.
\newblock \bibinfo{title}{Q3 Earnings Call: {{CEO}}'s Remarks}.
\newblock \bibinfo{howpublished}{https://blog.google/inside-google/message-ceo/alphabet-earnings-q3-2024/}.
\newblock


\bibitem[Ajami et~al\mbox{.}(2019)]%
        {Ajami_Woodbridge_Feitelson_2019}
\bibfield{author}{\bibinfo{person}{Shulamyt Ajami}, \bibinfo{person}{Yonatan Woodbridge}, {and} \bibinfo{person}{Dror~G. Feitelson}.} \bibinfo{year}{2019}\natexlab{}.
\newblock \showarticletitle{Syntax, predicates, idioms — what really affects code complexity?}
\newblock \bibinfo{journal}{\emph{Empirical Software Engineering}} \bibinfo{volume}{24}, \bibinfo{number}{1} (\bibinfo{date}{Feb.} \bibinfo{year}{2019}), \bibinfo{pages}{287–328}.
\newblock
\showISSN{1573-7616}
\urldef\tempurl%
\url{https://doi.org/10.1007/s10664-018-9628-3}
\showDOI{\tempurl}


\bibitem[Avidan and Feitelson(2017)]%
        {Avidan_2017}
\bibfield{author}{\bibinfo{person}{Eran Avidan} {and} \bibinfo{person}{Dror~G. Feitelson}.} \bibinfo{year}{2017}\natexlab{}.
\newblock \showarticletitle{Effects of Variable Names on Comprehension: An Empirical Study}. In \bibinfo{booktitle}{\emph{2017 IEEE/ACM 25th International Conference on Program Comprehension (ICPC)}}. \bibinfo{pages}{55–65}.
\newblock
\urldef\tempurl%
\url{https://doi.org/10.1109/ICPC.2017.27}
\showDOI{\tempurl}


\bibitem[Babe et~al\mbox{.}(2023)]%
        {babe_StudentEvalBenchmarkStudentWritten_2023}
\bibfield{author}{\bibinfo{person}{Hannah~McLean Babe}, \bibinfo{person}{Sydney Nguyen}, \bibinfo{person}{Yangtian Zi}, \bibinfo{person}{Arjun Guha}, \bibinfo{person}{Molly~Q. Feldman}, {and} \bibinfo{person}{Carolyn~Jane Anderson}.} \bibinfo{year}{2023}\natexlab{}.
\newblock \bibinfo{title}{{{StudentEval}}: {{A Benchmark}} of {{Student-Written Prompts}} for {{Large Language Models}} of {{Code}}}.
\newblock
\newblock
\urldef\tempurl%
\url{https://doi.org/10.48550/arXiv.2306.04556}
\showDOI{\tempurl}
\showeprint[arxiv]{2306.04556}~[cs]


\bibitem[Bauer et~al\mbox{.}(2019)]%
        {Bauer_2019}
\bibfield{author}{\bibinfo{person}{Jennifer Bauer}, \bibinfo{person}{Janet Siegmund}, \bibinfo{person}{Norman Peitek}, \bibinfo{person}{Johannes~C. Hofmeister}, {and} \bibinfo{person}{Sven Apel}.} \bibinfo{year}{2019}\natexlab{}.
\newblock \showarticletitle{Indentation: Simply a Matter of Style or Support for Program Comprehension?}. In \bibinfo{booktitle}{\emph{2019 IEEE/ACM 27th International Conference on Program Comprehension (ICPC)}}. \bibinfo{pages}{154–164}.
\newblock
\showISSN{2643-7171}
\urldef\tempurl%
\url{https://doi.org/10.1109/ICPC.2019.00033}
\showDOI{\tempurl}


\bibitem[Bednarik and Tukiainen(2006)]%
        {Bednarik_2006}
\bibfield{author}{\bibinfo{person}{Roman Bednarik} {and} \bibinfo{person}{Markku Tukiainen}.} \bibinfo{year}{2006}\natexlab{}.
\newblock \showarticletitle{An eye-tracking methodology for characterizing program comprehension processes}. In \bibinfo{booktitle}{\emph{Proceedings of the 2006 symposium on Eye tracking research \& applications}} \emph{(\bibinfo{series}{ETRA ’06})}. \bibinfo{publisher}{Association for Computing Machinery}, \bibinfo{address}{New York, NY, USA}, \bibinfo{pages}{125–132}.
\newblock
\showISBNx{978-1-59593-305-8}
\urldef\tempurl%
\url{https://doi.org/10.1145/1117309.1117356}
\showDOI{\tempurl}


\bibitem[Carrington(1997)]%
        {carrington_TeachingSoftwareTesting_1997}
\bibfield{author}{\bibinfo{person}{David Carrington}.} \bibinfo{year}{1997}\natexlab{}.
\newblock \showarticletitle{Teaching Software Testing}. In \bibinfo{booktitle}{\emph{Proceedings of the 2nd {{Australasian}} Conference on {{Computer}} Science Education}} \emph{(\bibinfo{series}{{{ACSE}} '97})}. \bibinfo{publisher}{Association for Computing Machinery}, \bibinfo{address}{New York, NY, USA}, \bibinfo{pages}{59--64}.
\newblock
\showISBNx{978-0-89791-958-6}
\urldef\tempurl%
\url{https://doi.org/10.1145/299359.299369}
\showDOI{\tempurl}


\bibitem[Castelhano et~al\mbox{.}(2019)]%
        {Castelhano_2019}
\bibfield{author}{\bibinfo{person}{Joao Castelhano}, \bibinfo{person}{Isabel~C. Duarte}, \bibinfo{person}{Carlos Ferreira}, \bibinfo{person}{Joao Duraes}, \bibinfo{person}{Henrique Madeira}, {and} \bibinfo{person}{Miguel Castelo-Branco}.} \bibinfo{year}{2019}\natexlab{}.
\newblock \showarticletitle{The role of the insula in intuitive expert bug detection in computer code: an fMRI study}.
\newblock \bibinfo{journal}{\emph{Brain Imaging and Behavior}} \bibinfo{volume}{13}, \bibinfo{number}{3} (\bibinfo{date}{June} \bibinfo{year}{2019}), \bibinfo{pages}{623–637}.
\newblock
\showISSN{1931-7565}
\urldef\tempurl%
\url{https://doi.org/10.1007/s11682-018-9885-1}
\showDOI{\tempurl}


\bibitem[Chen et~al\mbox{.}(2021)]%
        {chen_valuatingLargeLanguage_2021}
\bibfield{author}{\bibinfo{person}{Mark Chen}, \bibinfo{person}{Jerry Tworek}, \bibinfo{person}{Heewoo Jun}, \bibinfo{person}{Qiming Yuan}, \bibinfo{person}{Henrique Ponde de~Oliveira Pinto}, \bibinfo{person}{Jared Kaplan}, \bibinfo{person}{Harri Edwards}, \bibinfo{person}{Yuri Burda}, \bibinfo{person}{Nicholas Joseph}, \bibinfo{person}{Greg Brockman}, \bibinfo{person}{Alex Ray}, \bibinfo{person}{Raul Puri}, \bibinfo{person}{Gretchen Krueger}, \bibinfo{person}{Michael Petrov}, \bibinfo{person}{Heidy Khlaaf}, \bibinfo{person}{Girish Sastry}, \bibinfo{person}{Pamela Mishkin}, \bibinfo{person}{Brooke Chan}, \bibinfo{person}{Scott Gray}, \bibinfo{person}{Nick Ryder}, \bibinfo{person}{Mikhail Pavlov}, \bibinfo{person}{Alethea Power}, \bibinfo{person}{Lukasz Kaiser}, \bibinfo{person}{Mohammad Bavarian}, \bibinfo{person}{Clemens Winter}, \bibinfo{person}{Philippe Tillet}, \bibinfo{person}{Felipe~Petroski Such}, \bibinfo{person}{Dave Cummings}, \bibinfo{person}{Matthias Plappert}, \bibinfo{person}{Fotios Chantzis},
  \bibinfo{person}{Elizabeth Barnes}, \bibinfo{person}{Ariel {Herbert-Voss}}, \bibinfo{person}{William~Hebgen Guss}, \bibinfo{person}{Alex Nichol}, \bibinfo{person}{Alex Paino}, \bibinfo{person}{Nikolas Tezak}, \bibinfo{person}{Jie Tang}, \bibinfo{person}{Igor Babuschkin}, \bibinfo{person}{Suchir Balaji}, \bibinfo{person}{Shantanu Jain}, \bibinfo{person}{William Saunders}, \bibinfo{person}{Christopher Hesse}, \bibinfo{person}{Andrew~N. Carr}, \bibinfo{person}{Jan Leike}, \bibinfo{person}{Josh Achiam}, \bibinfo{person}{Vedant Misra}, \bibinfo{person}{Evan Morikawa}, \bibinfo{person}{Alec Radford}, \bibinfo{person}{Matthew Knight}, \bibinfo{person}{Miles Brundage}, \bibinfo{person}{Mira Murati}, \bibinfo{person}{Katie Mayer}, \bibinfo{person}{Peter Welinder}, \bibinfo{person}{Bob McGrew}, \bibinfo{person}{Dario Amodei}, \bibinfo{person}{Sam McCandlish}, \bibinfo{person}{Ilya Sutskever}, {and} \bibinfo{person}{Wojciech Zaremba}.} \bibinfo{year}{2021}\natexlab{}.
\newblock \bibinfo{title}{Evaluating {{Large Language Models Trained}} on {{Code}}}.
\newblock
\newblock
\showeprint[arxiv]{2107.03374}~[cs]


\bibitem[De-Arteaga et~al\mbox{.}(2020)]%
        {de-arteaga_case_2020}
\bibfield{author}{\bibinfo{person}{Maria De-Arteaga}, \bibinfo{person}{Riccardo Fogliato}, {and} \bibinfo{person}{Alexandra Chouldechova}.} \bibinfo{year}{2020}\natexlab{}.
\newblock \showarticletitle{A {Case} for {Humans}-in-the-{Loop}: {Decisions} in the {Presence} of {Erroneous} {Algorithmic} {Scores}}. In \bibinfo{booktitle}{\emph{Proceedings of the 2020 {CHI} {Conference} on {Human} {Factors} in {Computing} {Systems}}} \emph{(\bibinfo{series}{{CHI} '20})}. \bibinfo{publisher}{Association for Computing Machinery}, \bibinfo{address}{New York, NY, USA}, \bibinfo{pages}{1--12}.
\newblock
\showISBNx{978-1-4503-6708-0}
\urldef\tempurl%
\url{https://doi.org/10.1145/3313831.3376638}
\showDOI{\tempurl}


\bibitem[Denny et~al\mbox{.}(2024a)]%
        {denny_PromptProblemsNew_2024}
\bibfield{author}{\bibinfo{person}{Paul Denny}, \bibinfo{person}{Juho Leinonen}, \bibinfo{person}{James Prather}, \bibinfo{person}{Andrew {Luxton-Reilly}}, \bibinfo{person}{Thezyrie Amarouche}, \bibinfo{person}{Brett~A. Becker}, {and} \bibinfo{person}{Brent~N. Reeves}.} \bibinfo{year}{2024}\natexlab{a}.
\newblock \showarticletitle{Prompt {{Problems}}: {{A New Programming Exercise}} for the {{Generative AI Era}}}. In \bibinfo{booktitle}{\emph{Proceedings of the 55th {{ACM Technical Symposium}} on {{Computer Science Education V}}. 1}} \emph{(\bibinfo{series}{{{SIGCSE}} 2024})}. \bibinfo{publisher}{Association for Computing Machinery}, \bibinfo{address}{New York, NY, USA}, \bibinfo{pages}{296--302}.
\newblock
\showISBNx{9798400704239}
\urldef\tempurl%
\url{https://doi.org/10.1145/3626252.3630909}
\showDOI{\tempurl}


\bibitem[Denny et~al\mbox{.}(2024b)]%
        {denny_ExplainingCodePurpose_2024}
\bibfield{author}{\bibinfo{person}{Paul Denny}, \bibinfo{person}{David~H. Smith}, \bibinfo{person}{Max Fowler}, \bibinfo{person}{James Prather}, \bibinfo{person}{Brett~A. Becker}, {and} \bibinfo{person}{Juho Leinonen}.} \bibinfo{year}{2024}\natexlab{b}.
\newblock \showarticletitle{Explaining {{Code}} with a {{Purpose}}: {{An Integrated Approach}} for {{Developing Code Comprehension}} and {{Prompting Skills}}}. In \bibinfo{booktitle}{\emph{Proceedings of the 2024 on {{Innovation}} and {{Technology}} in {{Computer Science Education V}}. 1}}. \bibinfo{publisher}{ACM}, \bibinfo{address}{Milan Italy}, \bibinfo{pages}{283--289}.
\newblock
\showISBNx{979-8-4007-0600-4}
\urldef\tempurl%
\url{https://doi.org/10.1145/3649217.3653587}
\showDOI{\tempurl}


\bibitem[Dibia et~al\mbox{.}(2023)]%
        {dibia_aligning_2023}
\bibfield{author}{\bibinfo{person}{Victor Dibia}, \bibinfo{person}{Adam Fourney}, \bibinfo{person}{Gagan Bansal}, \bibinfo{person}{Forough Poursabzi-Sangdeh}, \bibinfo{person}{Han Liu}, {and} \bibinfo{person}{Saleema Amershi}.} \bibinfo{year}{2023}\natexlab{}.
\newblock \bibinfo{title}{Aligning Offline Metrics and Human Judgments of Value for Code Generation Models}.
\newblock , \bibinfo{numpages}{8516--8528}~pages.
\newblock
\urldef\tempurl%
\url{https://doi.org/10.18653/v1/2023.findings-acl.540}
\showDOI{\tempurl}


\bibitem[Edwards(2003)]%
        {edwards_ImprovingStudentPerformance_2003}
\bibfield{author}{\bibinfo{person}{Stephen~H. Edwards}.} \bibinfo{year}{2003}\natexlab{}.
\newblock \showarticletitle{Improving Student Performance by Evaluating How Well Students Test Their Own Programs}.
\newblock \bibinfo{journal}{\emph{J. Educ. Resour. Comput.}} \bibinfo{volume}{3}, \bibinfo{number}{3} (\bibinfo{date}{Sept.} \bibinfo{year}{2003}), \bibinfo{pages}{1--es}.
\newblock
\showISSN{1531-4278}
\urldef\tempurl%
\url{https://doi.org/10.1145/1029994.1029995}
\showDOI{\tempurl}


\bibitem[Feldman and Anderson(2024)]%
        {feldman_non-expert_2024}
\bibfield{author}{\bibinfo{person}{Molly~Q Feldman} {and} \bibinfo{person}{Carolyn~Jane Anderson}.} \bibinfo{year}{2024}\natexlab{}.
\newblock \showarticletitle{Non-{Expert} {Programmers} in the {Generative} {AI} {Future}}. In \bibinfo{booktitle}{\emph{Proceedings of the 3rd {Annual} {Meeting} of the {Symposium} on {Human}-{Computer} {Interaction} for {Work}}}. \bibinfo{publisher}{ACM}, \bibinfo{address}{Newcastle upon Tyne United Kingdom}, \bibinfo{pages}{1--19}.
\newblock
\showISBNx{9798400710179}
\urldef\tempurl%
\url{https://doi.org/10.1145/3663384.3663393}
\showDOI{\tempurl}


\bibitem[Felleisen et~al\mbox{.}(2018)]%
        {felleisen2018design}
\bibfield{author}{\bibinfo{person}{Matthias Felleisen}, \bibinfo{person}{Robert~Bruce Findler}, \bibinfo{person}{Matthew Flatt}, {and} \bibinfo{person}{Shriram Krishnamurthi}.} \bibinfo{year}{2018}\natexlab{}.
\newblock \bibinfo{booktitle}{\emph{How to design programs: an introduction to programming and computing}}.
\newblock \bibinfo{publisher}{MIT Press}.
\newblock


\bibitem[Feng et~al\mbox{.}(2024)]%
        {feng_CourseAssistPedagogicallyAppropriate_2024}
\bibfield{author}{\bibinfo{person}{Ty Feng}, \bibinfo{person}{Sa Liu}, {and} \bibinfo{person}{Dipak Ghosal}.} \bibinfo{year}{2024}\natexlab{}.
\newblock \showarticletitle{CourseAssist: Pedagogically Appropriate AI Tutor for Computer Science Education}. In \bibinfo{booktitle}{\emph{Proceedings of the 2024 on ACM Virtual Global Computing Education Conference V. 2}} (Virtual Event, NC, USA) \emph{(\bibinfo{series}{SIGCSE Virtual 2024})}. \bibinfo{publisher}{Association for Computing Machinery}, \bibinfo{address}{New York, NY, USA}, \bibinfo{pages}{310–311}.
\newblock
\showISBNx{9798400706042}
\urldef\tempurl%
\url{https://doi.org/10.1145/3649409.3691094}
\showDOI{\tempurl}


\bibitem[for Women \& Information~Technology(2023)]%
        {noauthor_ncwit_nodate}
\bibfield{author}{\bibinfo{person}{National~Center for Women \& Information~Technology}.} \bibinfo{year}{2023}\natexlab{}.
\newblock \bibinfo{title}{NCWIT Guide to Demographic Survey Questions}.
\newblock
\newblock
\urldef\tempurl%
\url{https://docs.google.com/document/d/1E_CSANwOqbKjEG27woNbGZ09JIXUfAf4Cp9j8g5DFak}
\showURL{%
\tempurl}


\bibitem[Gadala(2017)]%
        {gadala_automation_2017}
\bibfield{author}{\bibinfo{person}{Marwa Gadala}.} \bibinfo{year}{2017}\natexlab{}.
\newblock \showarticletitle{Automation bias: exploring causal mechanisms and potential mitigation strategies}.
\newblock
\urldef\tempurl%
\url{https://api.semanticscholar.org/CorpusID:41123263}
\showURL{%
\tempurl}


\bibitem[Goddard et~al\mbox{.}(2012)]%
        {goddard_automation_2012}
\bibfield{author}{\bibinfo{person}{Kate Goddard}, \bibinfo{person}{Abdul~V. Roudsari}, {and} \bibinfo{person}{Jeremy~C. Wyatt}.} \bibinfo{year}{2012}\natexlab{}.
\newblock \showarticletitle{Automation bias: a systematic review of frequency, effect mediators, and mitigators}.
\newblock \bibinfo{journal}{\emph{Journal of the American Medical Informatics Association : JAMIA}}  \bibinfo{volume}{19 1} (\bibinfo{year}{2012}), \bibinfo{pages}{121--7}.
\newblock


\bibitem[Hart and Staveland(1988)]%
        {hart_development_1988}
\bibfield{author}{\bibinfo{person}{Sandra~G. Hart} {and} \bibinfo{person}{Lowell~E. Staveland}.} \bibinfo{year}{1988}\natexlab{}.
\newblock \showarticletitle{Development of {NASA}-{TLX} ({Task} {Load} {Index}): {Results} of {Empirical} and {Theoretical} {Research}}.
\newblock In \bibinfo{booktitle}{\emph{Advances in {Psychology}}}, \bibfield{editor}{\bibinfo{person}{Peter~A. Hancock} {and} \bibinfo{person}{Najmedin Meshkati}} (Eds.). \bibinfo{series}{Human {Mental} {Workload}}, Vol.~\bibinfo{volume}{52}. \bibinfo{publisher}{North-Holland}, \bibinfo{pages}{139--183}.
\newblock
\urldef\tempurl%
\url{https://doi.org/10.1016/S0166-4115(08)62386-9}
\showDOI{\tempurl}


\bibitem[Kazemitabaar et~al\mbox{.}(2024)]%
        {kazemitabaar_CodeAidEvaluatingClassroom_2024}
\bibfield{author}{\bibinfo{person}{Majeed Kazemitabaar}, \bibinfo{person}{Runlong Ye}, \bibinfo{person}{Xiaoning Wang}, \bibinfo{person}{Austin~Zachary Henley}, \bibinfo{person}{Paul Denny}, \bibinfo{person}{Michelle Craig}, {and} \bibinfo{person}{Tovi Grossman}.} \bibinfo{year}{2024}\natexlab{}.
\newblock \showarticletitle{{{CodeAid}}: {{Evaluating}} a {{Classroom Deployment}} of an {{LLM-based Programming Assistant}} That {{Balances Student}} and {{Educator Needs}}}. In \bibinfo{booktitle}{\emph{Proceedings of the {{CHI Conference}} on {{Human Factors}} in {{Computing Systems}}}}. \bibinfo{publisher}{ACM}, \bibinfo{address}{Honolulu HI USA}, \bibinfo{pages}{1--20}.
\newblock
\showISBNx{979-8-4007-0330-0}
\urldef\tempurl%
\url{https://doi.org/10.1145/3613904.3642773}
\showDOI{\tempurl}


\bibitem[{Kyle Daigle, GitHub Staff}(2024)]%
        {kyledaiglegithubstaffSurveyAIWave2024}
\bibfield{author}{\bibinfo{person}{{Kyle Daigle, GitHub Staff}}.} \bibinfo{year}{2024}\natexlab{}.
\newblock \bibinfo{title}{Survey: {{The AI}} Wave Continues to Grow on Software Development Teams}.
\newblock
\newblock
\urldef\tempurl%
\url{https://github.blog/news-insights/research/survey-ai-wave-grows}
\showURL{%
\tempurl}


\bibitem[Li et~al\mbox{.}(2024)]%
        {li_LeveragingLargeLanguage_2024}
\bibfield{author}{\bibinfo{person}{Zhen Li}, \bibinfo{person}{Xiaohan Xu}, \bibinfo{person}{Tao Shen}, \bibinfo{person}{Can Xu}, \bibinfo{person}{Jia-Chen Gu}, \bibinfo{person}{Yuxuan Lai}, \bibinfo{person}{Chongyang Tao}, {and} \bibinfo{person}{Shuai Ma}.} \bibinfo{year}{2024}\natexlab{}.
\newblock \showarticletitle{Leveraging {{Large Language Models}} for {{NLG Evaluation}}: {{Advances}} and {{Challenges}}}. In \bibinfo{booktitle}{\emph{Proceedings of the 2024 {{Conference}} on {{Empirical Methods}} in {{Natural Language Processing}}}}. \bibinfo{publisher}{Association for Computational Linguistics}, \bibinfo{address}{Miami, Florida, USA}, \bibinfo{pages}{16028--16045}.
\newblock
\urldef\tempurl%
\url{https://doi.org/10.18653/v1/2024.emnlp-main.896}
\showDOI{\tempurl}


\bibitem[Liu et~al\mbox{.}(2024)]%
        {liu_TeachingCS50AI_2024a}
\bibfield{author}{\bibinfo{person}{Rongxin Liu}, \bibinfo{person}{Carter Zenke}, \bibinfo{person}{Charlie Liu}, \bibinfo{person}{Andrew Holmes}, \bibinfo{person}{Patrick Thornton}, {and} \bibinfo{person}{David~J. Malan}.} \bibinfo{year}{2024}\natexlab{}.
\newblock \showarticletitle{Teaching {{CS50}} with {{AI}}: {{Leveraging Generative Artificial Intelligence}} in {{Computer Science Education}}}. In \bibinfo{booktitle}{\emph{Proceedings of the 55th {{ACM Technical Symposium}} on {{Computer Science Education V}}. 1}}. \bibinfo{publisher}{ACM}, \bibinfo{address}{Portland OR USA}, \bibinfo{pages}{750--756}.
\newblock
\showISBNx{9798400704239}
\urldef\tempurl%
\url{https://doi.org/10.1145/3626252.3630938}
\showDOI{\tempurl}


\bibitem[Lucchetti et~al\mbox{.}(2025)]%
        {lucchetti_SubstanceBeatsStyle_2024}
\bibfield{author}{\bibinfo{person}{Francesca Lucchetti}, \bibinfo{person}{Zixuan Wu}, \bibinfo{person}{Arjun Guha}, \bibinfo{person}{Molly~Q. Feldman}, {and} \bibinfo{person}{Carolyn~Jane Anderson}.} \bibinfo{year}{2025}\natexlab{}.
\newblock \showarticletitle{Substance {{Beats Style}}: {{Why Beginning Students Fail}} to {{Code}} with {{LLMs}}}. In \bibinfo{booktitle}{\emph{Proceedings of the 2025 Conference of the North American Chapter of the Association for Computational Linguistics: Human Language Technologies, Volume 1 (Long Papers)}}. \bibinfo{publisher}{Association for Computational Linguistics}, \bibinfo{address}{Albuquerque, USA}.
\newblock
\urldef\tempurl%
\url{https://doi.org/10.48550/arXiv.2410.19792}
\showDOI{\tempurl}
\newblock
\shownote{To appear}.


\bibitem[Ma et~al\mbox{.}(2024)]%
        {ma_WhatShouldWe_2024}
\bibfield{author}{\bibinfo{person}{Qianou Ma}, \bibinfo{person}{Weirui Peng}, \bibinfo{person}{Chenyang Yang}, \bibinfo{person}{Hua Shen}, \bibinfo{person}{Kenneth Koedinger}, {and} \bibinfo{person}{Tongshuang Wu}.} \bibinfo{year}{2024}\natexlab{}.
\newblock \bibinfo{title}{What {{Should We Engineer}} in {{Prompts}}? {{Training Humans}} in {{Requirement-Driven LLM Use}}}.
\newblock
\newblock
\urldef\tempurl%
\url{https://doi.org/10.48550/arXiv.2409.08775}
\showDOI{\tempurl}
\showeprint[arxiv]{2409.08775}~[cs]


\bibitem[McChesney and Bond(2019)]%
        {McChesney_2019}
\bibfield{author}{\bibinfo{person}{Ian McChesney} {and} \bibinfo{person}{Raymond Bond}.} \bibinfo{year}{2019}\natexlab{}.
\newblock \showarticletitle{Eye tracking analysis of computer program comprehension in programmers with dyslexia}.
\newblock \bibinfo{journal}{\emph{Empirical Software Engineering}} \bibinfo{volume}{24}, \bibinfo{number}{3} (\bibinfo{date}{June} \bibinfo{year}{2019}), \bibinfo{pages}{1109–1154}.
\newblock
\showISSN{1573-7616}
\urldef\tempurl%
\url{https://doi.org/10.1007/s10664-018-9649-y}
\showDOI{\tempurl}


\bibitem[Molina et~al\mbox{.}(2024)]%
        {molina_LeveragingLLMTutoring_2024}
\bibfield{author}{\bibinfo{person}{Ismael~Villegas Molina}, \bibinfo{person}{Audria Montalvo}, \bibinfo{person}{Benjamin Ochoa}, \bibinfo{person}{Paul Denny}, {and} \bibinfo{person}{Leo Porter}.} \bibinfo{year}{2024}\natexlab{}.
\newblock \bibinfo{title}{Leveraging {{LLM Tutoring Systems}} for {{Non-Native English Speakers}} in {{Introductory CS Courses}}}.
\newblock
\newblock
\urldef\tempurl%
\url{https://doi.org/10.48550/arXiv.2411.02725}
\showDOI{\tempurl}
\showeprint[arxiv]{2411.02725}~[cs]


\bibitem[Nelson et~al\mbox{.}(2017)]%
        {nelson_xie_ko_2017}
\bibfield{author}{\bibinfo{person}{Greg~L. Nelson}, \bibinfo{person}{Benjamin Xie}, {and} \bibinfo{person}{Amy~J. Ko}.} \bibinfo{year}{2017}\natexlab{}.
\newblock \showarticletitle{Comprehension First: Evaluating a Novel Pedagogy and Tutoring System for Program Tracing in CS1}. In \bibinfo{booktitle}{\emph{Proceedings of the 2017 ACM Conference on International Computing Education Research}}. \bibinfo{publisher}{ACM}, \bibinfo{address}{Tacoma Washington USA}, \bibinfo{pages}{2–11}.
\newblock
\showISBNx{978-1-4503-4968-0}
\urldef\tempurl%
\url{https://doi.org/10.1145/3105726.3106178}
\showDOI{\tempurl}


\bibitem[Nguyen et~al\mbox{.}(2024)]%
        {charlie}
\bibfield{author}{\bibinfo{person}{Sydney Nguyen}, \bibinfo{person}{Hannah~McLean Babe}, \bibinfo{person}{Yangtian Zi}, \bibinfo{person}{Arjun Guha}, \bibinfo{person}{Carolyn~Jane Anderson}, {and} \bibinfo{person}{Molly~Q Feldman}.} \bibinfo{year}{2024}\natexlab{}.
\newblock \showarticletitle{How {Beginning} {Programmers} and {Code} {LLMs} ({Mis})read {Each} {Other}}. In \bibinfo{booktitle}{\emph{Proceedings of the {CHI} {Conference} on {Human} {Factors} in {Computing} {Systems}}}. \bibinfo{publisher}{ACM}, \bibinfo{address}{Honolulu HI USA}, \bibinfo{pages}{1--26}.
\newblock
\showISBNx{9798400703300}
\urldef\tempurl%
\url{https://doi.org/10.1145/3613904.3642706}
\showDOI{\tempurl}


\bibitem[Peitek et~al\mbox{.}(2018)]%
        {Peitek_2018}
\bibfield{author}{\bibinfo{person}{Norman Peitek}, \bibinfo{person}{Janet Siegmund}, \bibinfo{person}{Chris Parnin}, \bibinfo{person}{Sven Apel}, \bibinfo{person}{Johannes~C. Hofmeister}, {and} \bibinfo{person}{André Brechmann}.} \bibinfo{year}{2018}\natexlab{}.
\newblock \showarticletitle{Simultaneous measurement of program comprehension with fMRI and eye tracking: a case study}. In \bibinfo{booktitle}{\emph{Proceedings of the 12th ACM/IEEE International Symposium on Empirical Software Engineering and Measurement}} \emph{(\bibinfo{series}{ESEM ’18})}. \bibinfo{publisher}{Association for Computing Machinery}, \bibinfo{address}{New York, NY, USA}, \bibinfo{pages}{1–10}.
\newblock
\showISBNx{978-1-4503-5823-1}
\urldef\tempurl%
\url{https://doi.org/10.1145/3239235.3240495}
\showDOI{\tempurl}


\bibitem[Prather et~al\mbox{.}(2024)]%
        {prather_widening_2024}
\bibfield{author}{\bibinfo{person}{James Prather}, \bibinfo{person}{Brent~N Reeves}, \bibinfo{person}{Juho Leinonen}, \bibinfo{person}{Stephen MacNeil}, \bibinfo{person}{Arisoa~S Randrianasolo}, \bibinfo{person}{Brett~A. Becker}, \bibinfo{person}{Bailey Kimmel}, \bibinfo{person}{Jared Wright}, {and} \bibinfo{person}{Ben Briggs}.} \bibinfo{year}{2024}\natexlab{}.
\newblock \showarticletitle{The {Widening} {Gap}: {The} {Benefits} and {Harms} of {Generative} {AI} for {Novice} {Programmers}}. In \bibinfo{booktitle}{\emph{Proceedings of the 2024 {ACM} {Conference} on {International} {Computing} {Education} {Research} - {Volume} 1}}. \bibinfo{publisher}{ACM}, \bibinfo{address}{Melbourne VIC Australia}, \bibinfo{pages}{469--486}.
\newblock
\showISBNx{9798400704758}
\urldef\tempurl%
\url{https://doi.org/10.1145/3632620.3671116}
\showDOI{\tempurl}


\bibitem[Rahe and Maalej(2025)]%
        {rahe_HowProgrammingStudents_2025}
\bibfield{author}{\bibinfo{person}{Christian Rahe} {and} \bibinfo{person}{Walid Maalej}.} \bibinfo{year}{2025}\natexlab{}.
\newblock \bibinfo{title}{How {{Do Programming Students Use Generative AI}}?}
\newblock
\newblock
\urldef\tempurl%
\url{https://doi.org/10.48550/arXiv.2501.10091}
\showDOI{\tempurl}
\showeprint[arxiv]{2501.10091}~[cs]


\bibitem[Rozi{\`e}re et~al\mbox{.}({[n.\,d.]})]%
        {roziere_CodeLlamaOpen}
\bibfield{author}{\bibinfo{person}{Baptiste Rozi{\`e}re}, \bibinfo{person}{Jonas Gehring}, \bibinfo{person}{Fabian Gloeckle}, \bibinfo{person}{Sten Sootla}, \bibinfo{person}{Itai Gat}, \bibinfo{person}{Ellen Tan}, \bibinfo{person}{Yossi Adi}, \bibinfo{person}{Jingyu Liu}, \bibinfo{person}{Romain Sauvestre}, \bibinfo{person}{Tal Remez}, \bibinfo{person}{J{\'e}r{\'e}my Rapin}, \bibinfo{person}{Ivan Evtimov}, \bibinfo{person}{Joanna Bitton}, \bibinfo{person}{Manish Bhatt}, \bibinfo{person}{Cristian~Canton Ferrer}, \bibinfo{person}{Wenhan Xiong}, \bibinfo{person}{Alexandre D{\'e}fossez}, \bibinfo{person}{Jade Copet}, \bibinfo{person}{Faisal Azhar}, \bibinfo{person}{Hugo Touvron}, \bibinfo{person}{Louis Martin}, \bibinfo{person}{Nicolas Usunier}, \bibinfo{person}{Thomas Scialom}, {and} \bibinfo{person}{Gabriel Synnaeve}.} \bibinfo{year}{[n.\,d.]}\natexlab{}.
\newblock \showarticletitle{Code {{Llama}}: {{Open Foundation Models}} for {{Code}}}.
\newblock  (\bibinfo{year}{[n.\,d.]}).
\newblock


\bibitem[Schroeder and {Wood-Doughty}(2024)]%
        {schroeder_CanYouTrust_2024}
\bibfield{author}{\bibinfo{person}{Kayla Schroeder} {and} \bibinfo{person}{Zach {Wood-Doughty}}.} \bibinfo{year}{2024}\natexlab{}.
\newblock \bibinfo{title}{Can {{You Trust LLM Judgments}}? {{Reliability}} of {{LLM-as-a-Judge}}}.
\newblock
\newblock
\urldef\tempurl%
\url{https://doi.org/10.48550/arXiv.2412.12509}
\showDOI{\tempurl}
\showeprint[arxiv]{2412.12509}~[cs]


\bibitem[Skita et~al\mbox{.}(2000)]%
        {skita_accountability_2000}
\bibfield{author}{\bibinfo{person}{Linda~J Skita}, \bibinfo{person}{Kathleen Mosier}, {and} \bibinfo{person}{Mark~D. Burdick}.} \bibinfo{year}{2000}\natexlab{}.
\newblock \showarticletitle{Accountability and automation bias}.
\newblock \bibinfo{journal}{\emph{International Journal of Human-Computer Studies}} \bibinfo{volume}{52}, \bibinfo{number}{4} (\bibinfo{year}{2000}), \bibinfo{pages}{701--717}.
\newblock
\showISSN{1071-5819}
\urldef\tempurl%
\url{https://doi.org/10.1006/ijhc.1999.0349}
\showDOI{\tempurl}


\bibitem[Tian et~al\mbox{.}(2025)]%
        {tian_CodeHaluInvestigatingCode_2025}
\bibfield{author}{\bibinfo{person}{Yuchen Tian}, \bibinfo{person}{Weixiang Yan}, \bibinfo{person}{Qian Yang}, \bibinfo{person}{Xuandong Zhao}, \bibinfo{person}{Qian Chen}, \bibinfo{person}{Wen Wang}, \bibinfo{person}{Ziyang Luo}, \bibinfo{person}{Lei Ma}, {and} \bibinfo{person}{Dawn Song}.} \bibinfo{year}{2025}\natexlab{}.
\newblock \bibinfo{title}{{{CodeHalu}}: {{Investigating Code Hallucinations}} in {{LLMs}} via {{Execution-based Verification}}}.
\newblock
\newblock
\urldef\tempurl%
\url{https://doi.org/10.48550/arXiv.2405.00253}
\showDOI{\tempurl}
\showeprint[arxiv]{2405.00253}~[cs]


\bibitem[Tyen et~al\mbox{.}(2024)]%
        {tyen_LLMsCannotFind_2024}
\bibfield{author}{\bibinfo{person}{Gladys Tyen}, \bibinfo{person}{Hassan Mansoor}, \bibinfo{person}{Victor Carbune}, \bibinfo{person}{Peter Chen}, {and} \bibinfo{person}{Tony Mak}.} \bibinfo{year}{2024}\natexlab{}.
\newblock \showarticletitle{{{LLMs}} Cannot Find Reasoning Errors, but Can Correct Them given the Error Location}. In \bibinfo{booktitle}{\emph{Findings of the {{Association}} for {{Computational Linguistics}}: {{ACL}} 2024}}, \bibfield{editor}{\bibinfo{person}{Lun-Wei Ku}, \bibinfo{person}{Andre Martins}, {and} \bibinfo{person}{Vivek Srikumar}} (Eds.). \bibinfo{publisher}{Association for Computational Linguistics}, \bibinfo{address}{Bangkok, Thailand}, \bibinfo{pages}{13894--13908}.
\newblock
\urldef\tempurl%
\url{https://doi.org/10.18653/v1/2024.findings-acl.826}
\showDOI{\tempurl}


\bibitem[Vadaparty et~al\mbox{.}(2024)]%
        {vadaparty_cs1-llm_2024}
\bibfield{author}{\bibinfo{person}{Annapurna Vadaparty}, \bibinfo{person}{Daniel Zingaro}, \bibinfo{person}{David~H. Smith~Iv}, \bibinfo{person}{Mounika Padala}, \bibinfo{person}{Christine Alvarado}, \bibinfo{person}{Jamie Gorson~Benario}, {and} \bibinfo{person}{Leo Porter}.} \bibinfo{year}{2024}\natexlab{}.
\newblock \showarticletitle{{CS1}-{LLM}: {Integrating} {LLMs} into {CS1} {Instruction}}. In \bibinfo{booktitle}{\emph{Proceedings of the 2024 on {Innovation} and {Technology} in {Computer} {Science} {Education} {V}. 1}}. \bibinfo{publisher}{ACM}, \bibinfo{address}{Milan Italy}, \bibinfo{pages}{297--303}.
\newblock
\showISBNx{9798400706004}
\urldef\tempurl%
\url{https://doi.org/10.1145/3649217.3653584}
\showDOI{\tempurl}


\bibitem[Wang et~al\mbox{.}(2024)]%
        {wang_AssessingFactualReliability_2024}
\bibfield{author}{\bibinfo{person}{Weixuan Wang}, \bibinfo{person}{Barry Haddow}, \bibinfo{person}{Alexandra Birch}, {and} \bibinfo{person}{Wei Peng}.} \bibinfo{year}{2024}\natexlab{}.
\newblock \showarticletitle{Assessing {{Factual Reliability}} of {{Large Language Model Knowledge}}}. In \bibinfo{booktitle}{\emph{Proceedings of the 2024 {{Conference}} of the {{North American Chapter}} of the {{Association}} for {{Computational Linguistics}}: {{Human Language Technologies}} ({{Volume}} 1: {{Long Papers}})}}, \bibfield{editor}{\bibinfo{person}{Kevin Duh}, \bibinfo{person}{Helena Gomez}, {and} \bibinfo{person}{Steven Bethard}} (Eds.). \bibinfo{publisher}{Association for Computational Linguistics}, \bibinfo{address}{Mexico City, Mexico}, \bibinfo{pages}{805--819}.
\newblock
\urldef\tempurl%
\url{https://doi.org/10.18653/v1/2024.naacl-long.46}
\showDOI{\tempurl}


\bibitem[{Zamfirescu-Pereira} et~al\mbox{.}(2024)]%
        {zamfirescu-pereira_61ABotReport_2024}
\bibfield{author}{\bibinfo{person}{J.~D. {Zamfirescu-Pereira}}, \bibinfo{person}{Laryn Qi}, \bibinfo{person}{Bj{\"o}rn Hartmann}, \bibinfo{person}{John DeNero}, {and} \bibinfo{person}{Narges Norouzi}.} \bibinfo{year}{2024}\natexlab{}.
\newblock \bibinfo{title}{{{61A Bot Report}}: {{AI Assistants}} in {{CS1 Save Students Homework Time}} and {{Reduce Demands}} on {{Staff}}. ({{Now What}}?)}.
\newblock
\newblock
\urldef\tempurl%
\url{https://doi.org/10.1145/3641554.3701864}
\showDOI{\tempurl}
\showeprint[arxiv]{2406.05600}~[cs]


\end{thebibliography}

\appendix

\section{Thematic Analysis}
\label{app:thematic-analysis}

For RQ3, we performed thematic analysis of free-response answers from the post-study survey and the ``What happened?'' fields completed at the end of each task. 
The first two authors individually reviewed all relevant responses, agreed on a set of themes, and then coded the entire dataset.
Since themes could appear in multiple contexts (e.g. across post-task and post-study survey responses), all responses were considered holistically when identifying and coding themes.
Themes were not mutually exclusive, as participants could report contrasting themes for different tasks.

The initial coding resulted in a Cohen's Kappa value of 0.77, indicating high inter-rater reliability.
Discrepancies in coding were subsequently resolved through discussion. 
Selected participant quotes, edited for grammar, spelling and anonymity, are presented with pseudonyms assigned to participants.
The complete codebook, along with the frequency of identified themes, are detailed in Table~\ref{tab:thematic-codes} and Table~\ref{tab:thematic-codes-pt2}.

\section{Additional Results}
\label{app:additional-results}

\subsubsection*{Sentiment of Challenge.}
Participants completed a subset of NASA TLX \cite{hart_development_1988} ratings in the post-study survey.
They found the task as mentally demanding as prompting a Code LLM \cite{charlie} with a rating of 3.9 (Table \ref{tab:nasatlx}). 
Students that performed worse in the code comprehension part generally find the task more mentally demanding (Kendell's, $\tau = -0.390$, $p = 0.006$).
They feel less successful (Kendell's, $\tau = -0.417$, $p = 0.004$) and have more negative sentiment towards the task (Kendell's, $\tau = -0.353$, $p = 0.010$).

\subsubsection*{Miscellaneous Observations.}
For 25/160 (15.6\%) tasks, participants attributed the failures to their own carelessness.
3 participants (9.4\%) reported making mistakes when entering task outputs because they interpreted the tasks as trick questions.
For example, \textsc{happyWanderer} encountered a task where the code would produce an error on all 3 inputs but only labeled 1 input as error-producing.
They later reflected, stating they ``\textit{thought it was a trick question}'' and that it ``\textit{felt wrong to put error for every single one.}''

\begin{table*}[ht]
    \centering
    \begin{tabular}{|p{0.5\linewidth}|p{0.35\linewidth}|p{0.08\linewidth}|}
        \hline
        \textbf{Question} & \textbf{Scale (1 to 7)} & \textbf{Mean} \\
        \hline
        How mentally demanding was the task? & Very low $\rightarrow$ Very high & 3.9 \\
        \hline
        How hurried or rushed was the pace of the task? & Very low $\rightarrow$ Very high & 2.7 \\
        \hline
        How successful were you? & Perfect $\rightarrow$ Failure & 4 \\
        \hline
        How insecure, stressed, or discouraged were you? & Very low $\rightarrow$ Very high & 3.4 \\
        \hline
    \end{tabular}
    \caption{Mean NASA-TLX \cite{hart_development_1988} Ratings.}
    \label{tab:nasatlx}
\end{table*}

\begin{table*}[ht]
    \centering
    \begin{tabular}{|p{0.3\linewidth}|p{0.05\linewidth}|p{0.1\linewidth}|p{0.1\linewidth}|p{0.1\linewidth}|p{0.12\linewidth}|}
    \hline
    \multirow{2}{\linewidth}{Self-reported Demographics Information} & \multirow{2}{*}{Count} & \multicolumn{4}{c|}{Success Rates} \\ \cline{3-6}
    & & Prompt & Code & Total & Prompt-Code $\Delta$ \\ \hline
    Uses Code GenAI daily                                    & 2                 & 0.67       & 0.30 & 0.48  & 0.37  \\
    Uses Code GenAI 2 - 4 times a week                        & 11                & 0.85       & 0.47 & 0.66  & 0.38  \\
    Uses Code GenAI once a week                               & 4                 & 0.83       & 0.53 & 0.68  & 0.30  \\
    Uses Code GenAI less than once a week, but at least once per month & 4          & 0.77       & 0.50 & 0.63  & 0.27  \\
    Uses Code GenAI occasionally                              & 3                 & 0.80       & 0.36 & 0.58  & 0.44  \\
    Does not use Code GenAI                                     & 8                 & 0.81       & 0.44 & 0.62  & 0.37  \\ \hline
    Gender: Female                                           & 20                & 0.82       & 0.47 & 0.64  & 0.35  \\
    Gender: Male                                             & 12                & 0.79       & 0.43 & 0.61  & 0.36  \\ \hline
    Race: Asian                                              & 15                & 0.76       & 0.45 & 0.60  & 0.31  \\
    Race: Black                                              & 2                 & 0.77       & 0.33 & 0.55  & 0.43  \\
    Race: White                                              & 12                & 0.87       & 0.43 & 0.65  & 0.44  \\
    Race: Other                                              & 3                 & 0.89       & 0.67 & 0.78  & 0.22  \\ \hline
    Raised multilingual including English               & 13                & \textbf{0.82} & 0.49 & 0.66  & 0.33  \\
    Raised with only English                & 12                & \textbf{0.88} & 0.43 & 0.66  & 0.44  \\
    Raised in other languages than English            & 7                 & \textbf{0.68} & 0.42 & 0.55  & 0.26  \\ \hline
    Computing Intensive Majors                                & 15                & 0.82       & 0.51 & 0.66  & 0.32  \\
    Other Majors                                             & 17                & 0.80       & 0.41 & 0.60  & 0.39  \\ \hline
    Prior Programming Experience                              & 12                & 0.82       & 0.48 & 0.65  & 0.34  \\
    No Prior Programming Experience                           & 20                & 0.81       & 0.44 & 0.62  & 0.37  \\ \hline
    First Generation University Student                       & 4                 & 0.77       & 0.48 & 0.62  & 0.28  \\
    Not First Generation University Student                   & 28                & 0.82       & 0.45 & 0.63  & 0.37  \\ \hline
    Attended Private High School                             & 12                & 0.79       & 0.44 & 0.61  & 0.35  \\
    Attended Public High School                              & 20                & 0.82       & 0.46 & 0.64  & 0.36  \\ \hline
    \end{tabular}
\caption{Success Rates of Participants by Demographic information. Bold numbers indicate statistical significance within the demographic grouping.}
\label{tab:demographic}
\end{table*}

\begin{table*}[ht]
\centering
\begin{tabular}{|m{0.11\textwidth}|m{0.25\textwidth}|m{0.45\textwidth}|m{0.05\textwidth}|}
\hline
\textbf{Category}                                   & \textbf{Thematic Code}                & \textbf{Definition} & \textbf{Count} \\ \hline
\multirow{12}{=}{Challenges with the task}
 & Human error                                 & A mistake made by and  reported by participant due to carelessness, etc.              & 19             \\ \cline{2-4} 
 & Mistake in code: inconsistency with prompt & A mistake made by and reported by participant due to the code being inconsistent with the prompt     & 10             \\ \cline{2-4} 
 & Mistake in code: others                    & Other mistakes made by and reported by participant             & 6              \\ \cline{2-4} 
 & Not understanding code: Python conventions & Participant reports not knowing Python conventions or other Pythonic style code, e.g. \texttt{[::-1]}              & 18             \\ \cline{2-4} 
 & Not understanding code: Standard library   & Participant reports not knowing a Python standard library function               & 13             \\ \cline{2-4} 
 & Not understanding code: Programming fundamentals & Participant demonstrates a lack of understanding of programming fundamentals (e.g. list index start at 0)           & 7              \\ \cline{2-4} 
 & Comments style: Lack of comments           & Participant reports too much comment hindering code understanding              & 4              \\ \cline{2-4} 
 & Comments style: Too many comments          & Participant reports too few comment hindering code understanding              & 5              \\ \cline{2-4} 
 & Unfamiliar code style                      & Participant reports the code style being unfamilliar to them              & 14             \\ \cline{2-4} 
 & Unclear/hard to follow prompt              & Participant reports that the prompt is too hard to follow             & 16             \\ \cline{2-4} 
 & Biased by prompt                             & Participant reports that they were biased by the problem as described by the prompt, assuming the code would work as described            & 3              \\ \cline{2-4} 
 & Interpreting tasks like a quiz/test        & Participant reports that the they are treating a task like a trick question in a quiz             & 3              \\ \hline
\multirow{4}{=}{Reason for not using GenAI} 
 & Concern of plagiarism                      & Participant doesn't use GenAI for code due to concerns of plagiarism              & 2              \\ \cline{2-4} 
 & Concern of AI over-reliance                 & Participant doesn't use GenAI for code due to concerns of AI over-reliance              & 4              \\ \cline{2-4} 
 & Distrust in AI correctness                 & Participant doesn't use GenAI for code due to distrust in AI correctness              & 1              \\ \cline{2-4} 
 & AI can’t fully understand user             & Participant doesn't use GenAI for code due to belief in AI misunderstanding user             & 1              \\ \hline
\multirow{4}{=}{Methodology to understand code}    
 & Rely on provided examples                  & Participant reports relying on provided examples              & 5              \\ \cline{2-4} 
 & Rely on own examples                       &  Participant reports relying on examples they come up with on their own              & 3              \\ \cline{2-4} 
 & Code tracing                               & Participant reports tracing code              & 19             \\ \cline{2-4} 
 & Inclination to familiar code               & Participant reports relying on code familiar to them              & 1              \\ \hline
\multirow{6}{=}{Ease of reading code}                    
 & Easy to Read                               & Participant reports that code is easy to read              & 23             \\ \cline{2-4} 
 & Hard to Read                               & Participant reports that code is hard to read              & 16             \\ \cline{2-4} 
 & Unfamiliar interface/colors                & Participant reports that the interface is unfamiliar to them              & 2              \\ \cline{2-4} 
 & Comments: lots                             & Participant mentions that code has a lot of comments              & 4              \\ \cline{2-4} 
 & Comments: few                           & Participant mentions that code has few comments              & 3              \\ \cline{2-4} 
 & Style of code                              & Participant commented about the style of code              & 9              \\ \hline
\end{tabular}
\caption{Thematic codes and counts with definitions, Part 1.}
\label{tab:thematic-codes}
\end{table*}

\begin{table*}[ht]
\centering
\begin{tabular}{|m{0.11\textwidth}|m{0.25\textwidth}|m{0.45\textwidth}|m{0.05\textwidth}|}
\hline
\textbf{Category}                                   & \textbf{Thematic Code}                & \textbf{Definition} & \textbf{Count} \\ \hline

\multirow{6}{=}{Task validity}                   
 & Helpful examples                           & Participant mentions examples being helpful              & 32             \\ \cline{2-4} 
 & Unhelpful examples                         & Participant mentions examples being unhelpful              & 0              \\ \cline{2-4} 
 & Helpful descriptions                       & Participant mentions descriptions (i.e. prompt) being helpful              & 13             \\ \cline{2-4} 
 & Unhelpful descriptions                     & Participant mentions descriptions being unhelpful              & 7              \\ \cline{2-4} 
 & Clear descriptions                         & Participant mentions descriptions being clear           & 26             \\ \cline{2-4} 
 & Unclear descriptions                       & Participant mentions descriptions being unclear              & 14             \\ \hline
\multirow{6}{=}{Miscellaneous}                  
 & Inaccurate descriptions                    & Participant mentions descriptions being inaccurate              & 8              \\ \cline{2-4} 
 & Inaccurate comments                        & Participant mentions comments being inaccurate              & 2              \\ \cline{2-4} 
 & Confirming naturalness                     & Participant mentions that code presented is natural to them              & 8              \\ \cline{2-4} 
 & Refuting naturalness                       & Participant mentions that code presented is not natural to them              & 18             \\ \cline{2-4} 
 & Found code novel                           & Participant mentions finding code is novel           & 1              \\ \hline
\end{tabular}
\caption{Thematic codes and counts with definitions, Part 2.}
\label{tab:thematic-codes-pt2}
\end{table*}

\end{document}